\documentclass{emulateapj}
\newcommand{\manq}{manqu\'{e}}
\newcommand{\msun}{M_\sun}
\usepackage{lscape}

\begin{document}
\title{Evolved Stars in the Core of the Massive Globular Cluster NGC
  2419\footnotemark[1]}
\footnotetext[1]{Based on observations with the NASA/ESA {\it Hubble Space
      Telescope},  
obtained at the Space Telescope Science Institute, which is operated
by the Association of Universities for Research in Astronomy, Inc.,
under NASA Contract NAS 5-26555.}
\author{Eric L. Sandquist, Jordan M. Hess}
\affil{Department of Astronomy, San Diego State University, 5500 Campanile
  Drive, San Diego, CA 92182}
\email{erics@sciences.sdsu.edu}

\begin{abstract}
  We present an analysis of optical and ultraviolet {\it Hubble Space
    Telescope} photometry for evolved stars in the core of the distant massive
  globular cluster NGC 2419.
  We characterize the horizontal branch (HB) population in detail including
  corrections for incompleteness on the long blue tail. The majority of the
  horizontal branch stars can be identified with two main groups (one slightly
  bluer than the instability strip, and the other at the extreme end of the
  HB).  We present a method for removing (to first order) lifetime effects
  from the distribution of HB stars to facilitate more accurate measurements
  of helium abundance for clusters with blue HBs and to clarify the
  distribution of stars reaching the zero-age HB. The population ratio $R =
  N_{HB} / N_{RGB}$ implies there may be slight helium enrichment among the
  EHB stars in the cluster, but that it is likely to be small ($\Delta Y <
  0.05$).  An examination of the upper main sequence does not reveal any sign
  of multiple populations indicative of helium enrichment.
  
  The stellar distribution allows us to follow how the two main types of stars
  evolve after the HB.  We find that the transition from stars that reach the
  asymptotic giant branch to stars that remain at high temperatures probably
  occurs among the extreme horizontal branch stars (EHB) at a larger
  temperature than predicted by canonical evolution models, but qualitatively
  consistent with helium-enriched models.  Through comparisons of optical
  CMDs, we present evidence that the EHB clump in NGC 2419 contains the end of
  the canonical horizontal branch, and that the boundary between the normal HB
  stars and blue hook stars shows up as a change in the density of stars in
  the CMD. This corresponds to a spectroscopically-verified gap in NGC 2808
  and an ``edge'' in $\omega$ Cen.  The more clearly visible HB gap at $V \sim
  23.5$ identified by \citet{rip} appears to be too bright.
    
  Once corrected for lifetime effects, we find that NGC 2419 is currently
  converting about 25 -- 31\% of the first-ascent red giant stars in its core
  into extreme blue horizontal branch stars --- the largest fraction for any
  known globular cluster.  A comparison of upper red giant branch with
  theoretical models indicates there is a slight deficiency of bright red
  giant stars.  This deficiency occurs far enough below the tip of the red
  giant branch that it is unlikely to be associated with the production of
  extreme horizontal branch stars via strong mass loss before the core helium
  flash.

\end{abstract}



\section{Introduction}

NGC 2419 is the fourth brightest globular cluster known in the Milky Way ($M_V
= -9.58$; \citealt{har96}).  NGC 2419's large mass would normally make it
ideal for studying short phases in the lives of evolved low-mass stars because
massive cluster (having more stars) should produce a greater chance of
catching {\it some} stars in those phases. Until recently though, it has been
neglected because it resides far in the outer halo of the Milky Way
(approximately 90 kpc from the center). Massive clusters can make it possible
to observe stars in short-lived evolution phases because there is a greater
chance of catching {\it some} star in such a phase when there are more
chances.

Evolved stars have unusually strong influences on the luminosity-integrated
properties of a galaxy: in old stellar populations, the overall color and the
ultraviolet emission result from them.  The luminous giants at the tip of the
red giant branch (TRGB) are frequently used as standard candles in resolved
stellar populations. Extremely blue HB (hereafter, EHB) stars are a leading
candidate for galactic ultraviolet emission.  These populations are linked at
least in an evolutionary sense, but it remains a longstanding problem to
explain the details of how HB stars (and more specifically the EHB stars) are
produced. There has been much recent interest in multiple stellar populations
having different chemical compositions (particularly helium, but also heavier
elements) within individual clusters (NGC 2808, \citealt{dan2808}; $\omega$
Cen, \citealt{bedomega}; M13, \citealt{cadm13}; NGC 6218, \citealt{cn6218};
NGC 6388, \citealt{busso}; NGC 6441, \citealt{cadn6441}) However, even
allowing for composition effects, it still seems to be necessary for giant
stars to lose most of the mass in their envelopes if EHB stars are to be
produced --- the turnoff mass of a helium-enriched population cannot be
reduced enough to have a giant star consume its envelope before it reaches the
TRGB.

In this paper, we examine deep photometry of the evolved stars in the core of
the cluster using high resolution {\it Hubble Space Telescope (HST)} imagery.
We have three main goals: to characterize brief evolutionary phases and their
effects on the integrated light of the cluster, to examine the paths (often
brief) that lead from one evolutionary phase to another, and to use the
stellar populations to constrain chemical evolution within the cluster. EHB
and bright red giant branch (RGB) stars are particularly important to the
integrated colors of clusters.  EHB stars can be seen in several earlier CMDs
for NGC 2419 \citep{har97,stetn2419}, although the impression was that the
population was small due to incompleteness at the faint end. \citet{dale}
recently presented a reduction of a portion of the available archived {\it
  HST} data for the cluster that clearly shows the faint end of the blue HB
tail, and that the EHB population is a substantial fraction of all HB stars in
the cluster. In section \ref{HB}, we conduct a deeper analysis of the HB
population of the cluster.

The details of the transition from the RGB to the HB remain poorly understood.
In the case of the massive cluster NGC 2808, \citet{sm07} found a
deficit of bright red giants compared to theoretical predictions that
may be linked to the EHB stars there --- if a star loses
enough mass on the red giant, it can leave the red giant branch before
the helium flash at the TRGB, igniting helium later in a ``hot flash''
that ultimately deposits it at the blue end of the HB. NGC 2419 appears to be
more efficient at producing EHB stars than most clusters, so in \S \ref{urgb}
we examine the upper RGB for signatures of EHB star production.

Integrated luminosity has been linked to a number of unusual groups of stars
within globular clusters, including blue stragglers (their relative frequency
is anti-correlated with total luminosity), extremely blue horizontal branch
stars, and populations with different age and/or chemical composition. To
date, NGC 2419 has shown little indication that it contains anything but a
stellar population with a single composition ([Fe/H] $= -2.1$; \citealt{sunt})
and age. However, NGC 2419 is now known to have a strongly double-peaked HB,
and so we examine the helium abundance indicator $R$ in \S \ref{R}.


\section{Observations and Data Reduction}\label{selstar}

\begin{figure}
\includegraphics[scale=0.5]{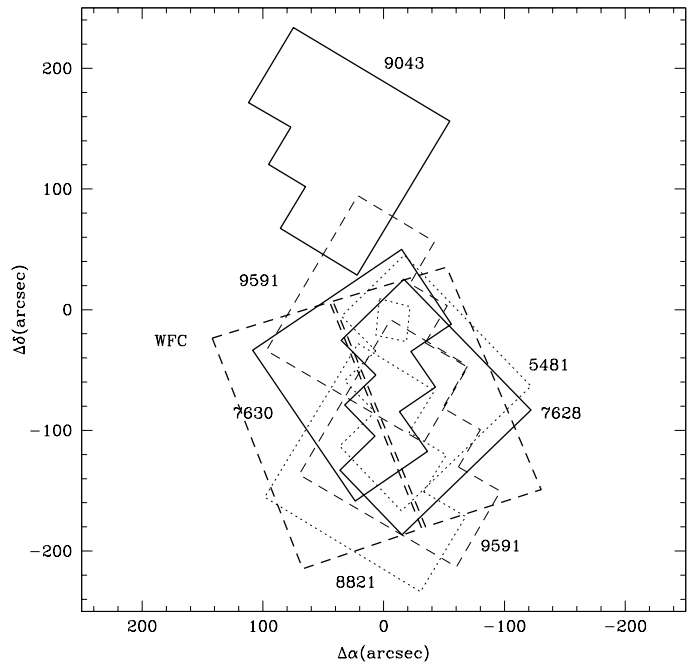} 
\caption{Sky positions of HST fields that were
    analyzed in this paper. Positions are given in arcseconds relative to the
    \citet{har96} position given for the cluster center. WFPC2 fields are
    identified with the proposal ID number (see Table \ref{sets}) except where
    the WFPC2 field almost completely overlaps another WFPC2 field. The deepest
    photometry was taken in the fields having bold lines. The ACS
    HRC field is the small rectangle surrounding (0,0).
    \label{pos}}
\end{figure}

This study has focused on the evolved stellar populations of the cluster core.
The archival HST datasets that were reduced are listed in Table \ref{sets},
and a map of the fields is shown in Fig. \ref{pos}.
Images from the ACS WFC and HRC instruments were processed using the DOLPHOT
photometry package\footnote{http://purcell.as.arizona.edu/~andy/dolphot/} with
its module tuned for ACS data.  Individual frames were obtained from the HST
archive in order to get photometry on the giant stars.  The WFPC2 images were
analyzed using the HSTPhot\footnote{http://purcell.as.arizona.edu/hstphot/}
photometry package \citep{hstphot}. Although different filters were used for
imaging in different epochs, the fields overlapped in many cases, and
in each field we have photometry in the F555W and F814W filters.

\begin{deluxetable*}{clcl}
\tablewidth{0pt} 
\tablecaption{HST Datasets for NGC 2419}
\tablehead{\colhead{Prop. ID\tablenotemark{a}} & \colhead{P.I.} & \colhead{Instrument} & \colhead{Filters}\label{sets}}
\startdata
5481 & Hesser & WFPC2 & F555W, F814W\\
\hline
7628 & Whitmore & WFPC2 & F336W, F380W, F439W, F450W, F555W, F675W, F814W \\ 
8095 & Ibata & WFPC2 & F555W \\
9601 & Koekemoer & WFPC2 & F300W, F336W, F439W, F450W, F675W \\
\hline
7630 & Casertano & WFPC2 & F300W, F555W, F814W\\
6937 & Stiavelli & WFPC2 & F555W\\
\hline
8821 & Riess & WFPC2 & F555W, F814W \\
\hline
9043 & Tonry & WFPC2 & F555W, F814W\\
\hline
9591 & Whitmore & WFPC2 & F814W\\
\hline
9666 & Gilliland & ACS WFC & F435W, F555W, F814W \\
\hline
10815& Brown & ACS HRC & F250W 
\enddata
\tablenotetext{a}{Observations with almost completely overlapping fields are
  grouped together.}
\end{deluxetable*}

An RA-DEC offset coordinate system was set up using the WFPC2 datasets.
Positions are given relative to the cluster center given by \citet{har96}.
This position agrees extremely well with the determination of the center of
gravity by \citet{dale}. Images in the ACS system are subject to a lot of
geometric distortion due to the off-axis positions of the cameras, so we used
fourth-order geometric transformations obtained from the most recent Image
Distortion Correction Table (IDCTAB) file
(qbu1641sj\_idc.fits)\footnote{http://www.stsci.edu/hst/observatory/cdbs/SIfileInfo/ACS/ACSDistortionCorrection}
in order to convert pixel positions into relative sky positions. Coordinates
for stars in the WFPC2 images were derived from the METRIC task in the IRAF
package STSDAS.  METRIC positions typically have absolute position errors that
are a significant fraction of an arcsecond, but we are primarily interested in
relative positions. After initial identifications of stars were made in the
different datasets, preliminary coordinate transformations were derived for
all of the WFPC2 and ACS fields.  The coordinate transformations were then
refined with additional cross-identifications, and these were used to match up
all stars measured in the datasets.

\begin{figure*}
  \includegraphics[scale=0.5,angle=-90]{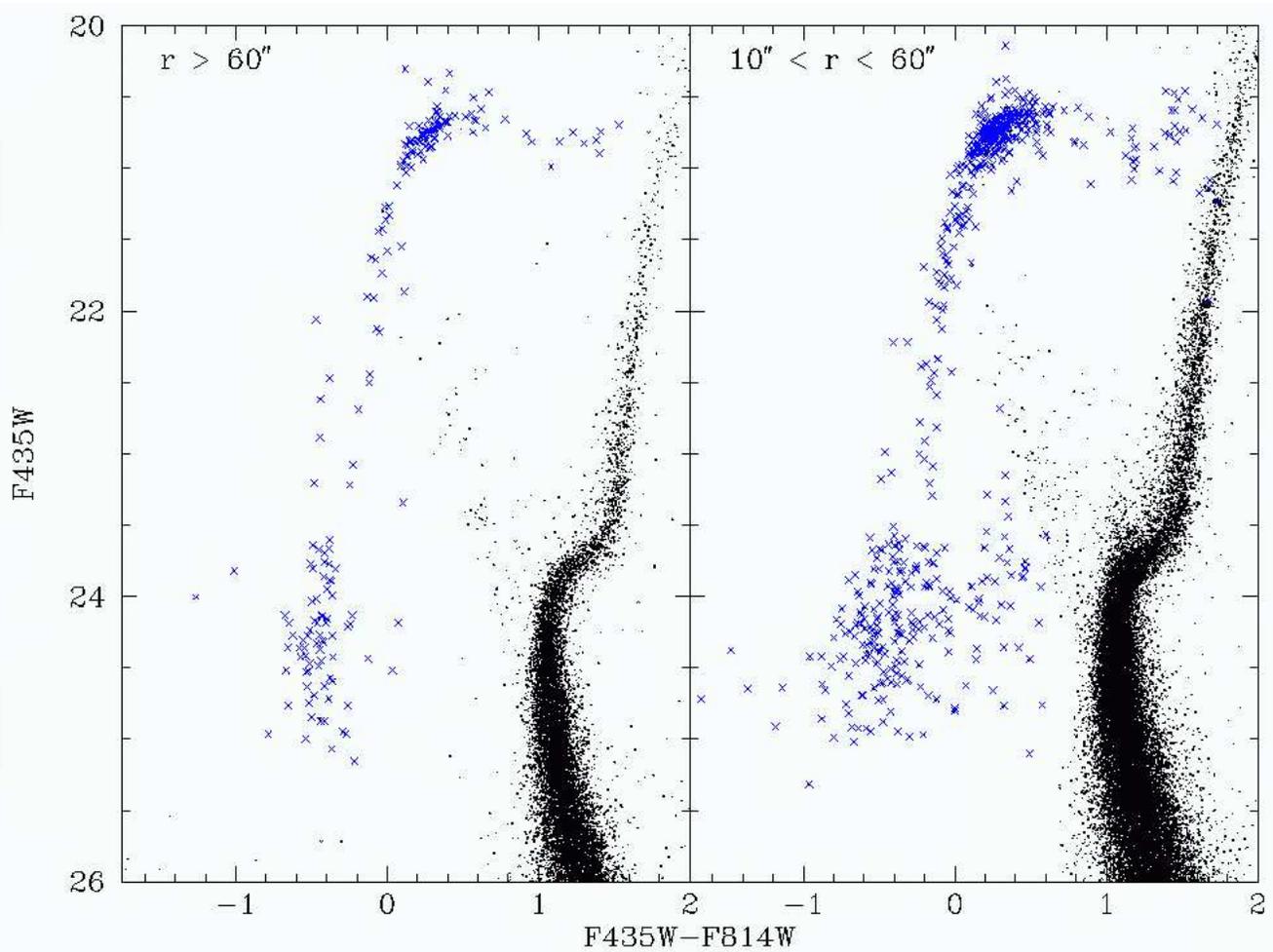} 
\caption{ACS WFC color-magnitude diagrams with horizontal
  branch stars passing criteria for detection in our artificial star trials
  ({\it blue crosses}; see \S \ref{artstar}), divided by projected distance
  from the cluster center. \label{hbr60}}
\end{figure*}

\begin{figure*}
  \includegraphics[scale=0.5,angle=-90]{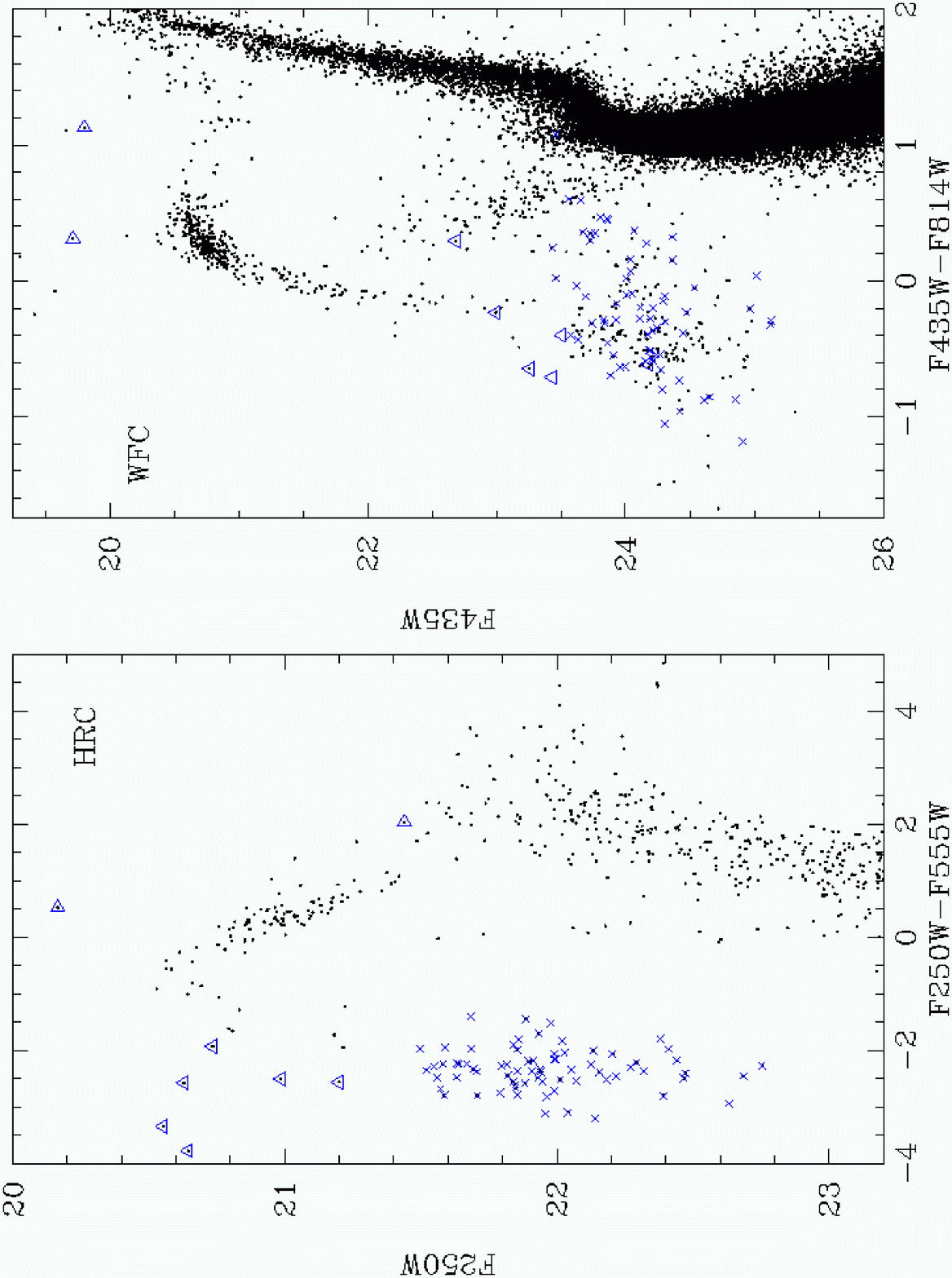} 
\caption{Color-magnitude diagrams showing the selection
    of extreme HB stars {\it (blue crosses)} from the ACS HRC F250W filter, and
    the corresponding stars in the ACS WFC field. Triangles show probable
    evolved HB stars in the ACS HRC field, with the triangles pointing 
 roughly in the
    direction of their expected evolution (see \S \ref{uvsec}).\label{f250}}
\end{figure*}

The HST datasets were analyzed to maximize the detection of different
populations of evolved stars in the cluster. Identification of horizontal
branch stars is generally more effective in shorter wavelength filters when
most of the HB stars are bluer than the RR Lyrae instability strip, as in NGC
2419. In this case, however, the optical CMDs from the ACS WFC field were deep
and had high spatial resolution. At the same time, the optical CMDs were
significantly incomplete in the core of the cluster even with the high
resolution. In Fig.  \ref{hbr60}, we show the optical WFC CMD for samples in
and outside $60\arcsec$ from the cluster center. From artificial star tests
(see \S \ref{artstar}), we find that the WFC data is more than 80\% complete
even down to the faint end of the EHB for $r > 60\arcsec$. Because they also
covered the widest area, they became our primary choice for selection. Had the
ultraviolet images been deeper, they would have been more reliable for
detecting the bluest HB stars. The ACS HRC frames in the F250W filter covered
a small field centered about $12\arcsec$ from the cluster core, allowing us to
confidently identify 11 faint HB stars that would otherwise have been rejected
by quality cuts that were applied to the next highest resolution dataset (the
ACS WFC images). Fig. \ref{f250} shows CMDs of HB stars detected in both the
HRC and WFC images. Based on these considerations, we believe the HB star
sample in the HRC field is virtually 100\% complete.  Observations in the
F336W filter of the proposal 7628 field also provide good detections of bright
EHB stars, although the detection limit brighter than the faint end of the EHB
(see Fig. \ref{m15f336}). The situation in the F300W filter was similar, so
they were only used for verification of brighter candidates.

Because some of the datasets have image sequences that are suitable for
searching for variables, we checked the RR Lyrae population in the HST fields.
We computed the $J$ statistic \citep{stetj} for the combination of F555W and
F814W images from the proposal ID 6937 and 7630 field, and found that this
efficiently identified variable stars in the field. We confirmed most of the
\citet{prrrlyr} variables that fall in the field (V2, V3, V4, V6, V11, V13,
V26, V29, V30, V33, V35, V38, V39, V40) and have detected others. The
coordinates given for V37 were inconsistent with the \citeauthor{prrrlyr}
finding chart position, which puts it outside the HST fields. The coordinates
for V38 and V40 were both significantly in error, but the stars were
identified using the finding chart and their variability was confirmed. We also
were unable to find an obvious counterpart to their V41. Most of the
RR Lyrae stars could be tentatively identified in the WFC datasets using
their CMD positions, but there were 4 stars that ended up in the middle of the
RGB as a result of the sampling of the light curve with a small number of
observations.

\begin{figure*}
  \includegraphics[scale=0.85]{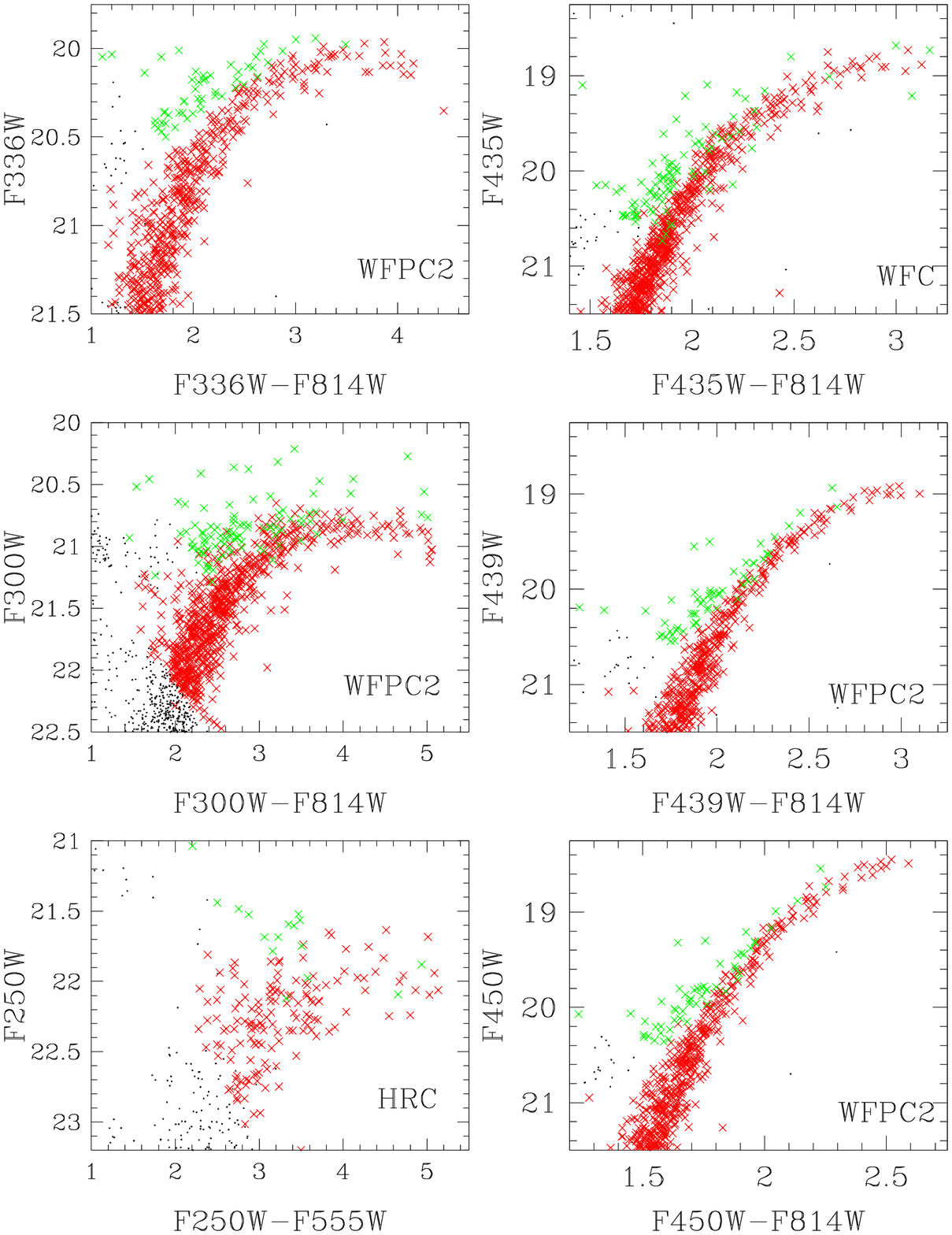}
 \caption{The color-magnitude diagrams used to
    select red giant branch branch {\it (red)} and asymptotic giant branch
    {\it (green)} stars in the HST fields. No cuts on
    position with the cluster were applied to any of the CMDs. \label{rgbsel}}
\end{figure*}

The CMDs used for the selection of RGB stars are shown in Fig. \ref{rgbsel}.
In the RGB star sample, the primary contaminant is AGB stars. For this reason
we also relied heavily on CMDs with short wavelength filters in order to
separate AGB and RGB stars by temperature. CMDs with ultraviolet observations
on the magnitude axis can help turn temperature (color) differences into
luminosity (magnitude) differences, which are often easier to detect. Our
primary choice for RGB identification used the F336W filter from the proposal
7628 field because this most cleanly and reliably separated AGB stars. As our
second choice we used ACS WFC observations in the F435W filter and WFPC2
observations in the F439W and F450W filters. Stars in the longest WFC
exposures were unfortunately saturated, although we were able to use saturated
star photometry from DOLPHOT in most cases. (We have not included saturated
star photometry in the averaged photometry described later.) If there was some
ambiguity in identifying a star as RGB or AGB, it usually came in the CMDs due
to (relatively) small color separation between RGB and AGB.  The AGB clump can
easily be seen in $19.9 < F435W < 20.5$, but the AGB sequence mostly merges
with the RGB brighter than that. Photometry in the F300W (WFPC2) and F250W
(HRC) filters were also useful, but again, we found that they were sometimes
unreliable toward the tip of the giant branch due to measurement
uncertainties. Toward the faint end of the sample though, there is a large and
useful magnitude difference between the RGB and the AGB clump.

\subsection{Artificial Star Tests}\label{artstar}

In order to fully characterize the faint HB population, we conducted a series
of artificial star tests on the ACS WFC frames, which comprises the great
majority of the surveyed field. Before we ran the tests, we first defined a
fiducial line running through the blue HB tail using the F435W filter for the
magnitude axis, and using both $F435W - F555W$ and $F435W - F814W$
colors. To ensure that the two artificial star colors were consistent, we used
the photometry for a select group of stars to define the fiducial. Artificial
star magnitudes were selected randomly with a flat luminosity function, and
colors were generated from quadratic interpolation between the fiducial
points.

In each of the artificial HB star tests, stars were placed in a grid of cells
30 pixels wide. The stars were randomly placed within cells, and the grid
itself was shifted from test to test. A total of 5 runs containing 91926 stars
were completed. Stars were identified as recovered if they were detected in
the F435W and F814W images and passed cuts on sharpness ($|\mbox{SHARP}| <
0.5$), $\chi^2$ ($< 5$), and crowding. With regard to crowding, the photometry
for a star is often affected by light from nearby neighbors. Photometry was
derived from profile fitting in which overlapping stellar images are fit
simultaneously with point-spread functions. The crowding measure involves the
resulting difference in brightness if nearby neighbors were {\it not}
simultaneously fitted. Stars were rejected if they would have been more than
0.3 mag brighter if nearby neighbors had been neglected. (While this probably
rejected some real stars, if we apply the cuts consistently to the artificial
stars and the real stars, we can derive accurate incompleteness corrections.)

\begin{figure}
  \includegraphics[scale=0.4]{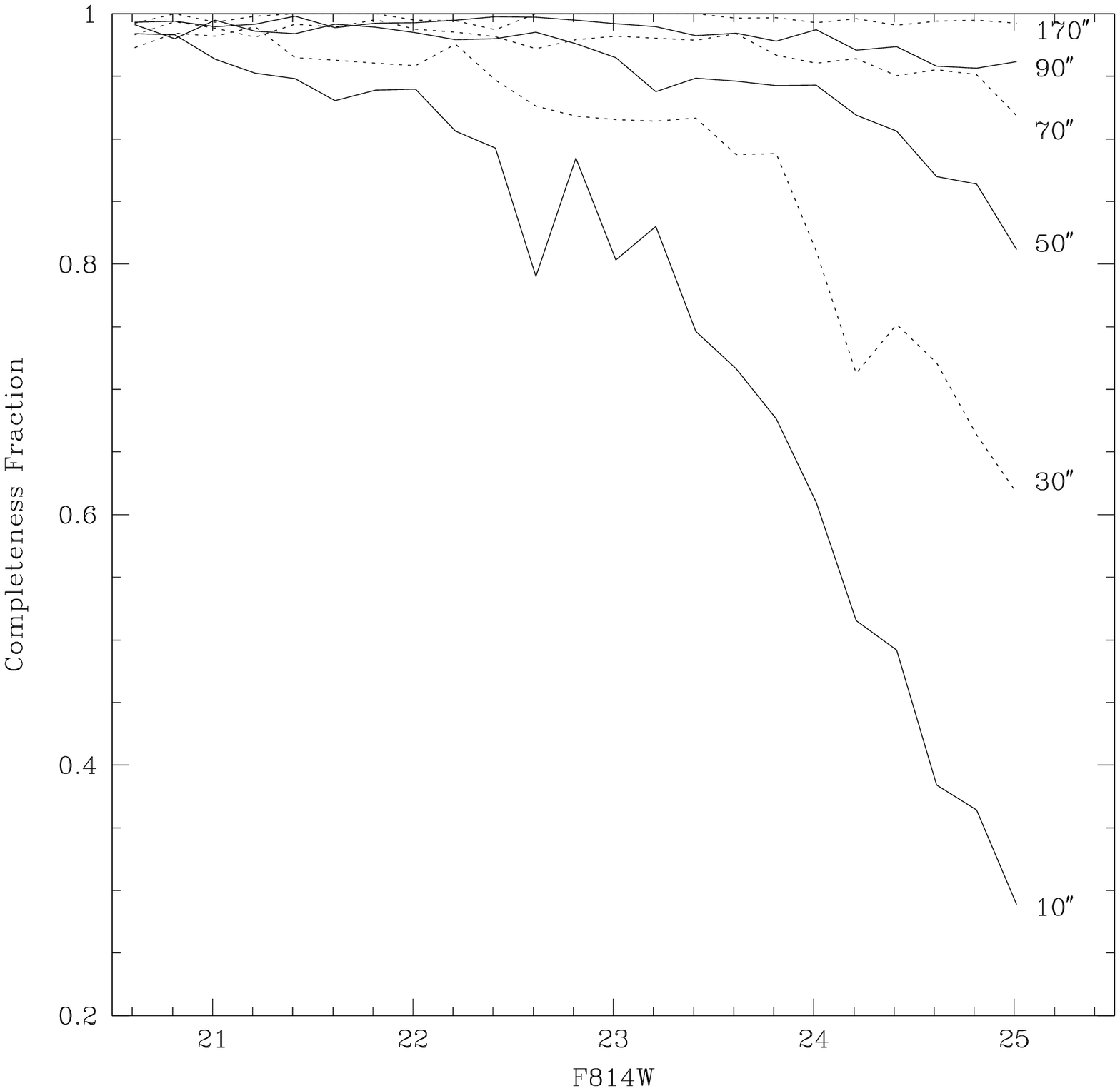} 
\caption{Completeness measures for artificial horizontal
    branch stars as a function of F814W magnitude and distance from the
    cluster center. The lines are identified with the radius of the center of
    the annulus (all annuli are $20\arcsec$ wide).    \label{incomp}}
\end{figure}

For the analysis, stars were divided into annuli ($20\arcsec$ wide;
approximately one core radius) and magnitude (0.2 mag wide) bins.  According
to the artificial star tests (see Fig. \ref{incomp}), the bright HB sample is
almost 100\% complete throughout the HST fields. The worst incompleteness
occurs in the cluster core, where EHB stars in the WFC field are only about
29\% complete within one core radius. However, we used UV observations to
improve the recovery of EHB stars in the densest part of the core. Because
crowding is negligible in the HRC images and the faint end of the HB is well
above the faint limit of the images, we have assumed that our HB sample is
100\% complete in the HRC field.

\subsection{Photometry}

With such a heterogeneous set of images and filters, we spent some effort to
merge the photometric datasets together where possible. We emphasize that the
merged photometry was {\it not} used in categorizing stars by evolutionary
state --- that was done using individual datasets because the relative
precision was higher.  We used the observations of the WFPC2 proposal 7630
field as the reference because it showed the greatest amount of overlap with
the other fields. (We found significant zeropoint differences between the two
ACS WFC chips, which is why they were not used.) The photometry that could be
merged in this way used the F300W, F555W, and F814W filters.

\begin{figure*}
\begin{center}
  \includegraphics[scale=0.5,angle=-90]{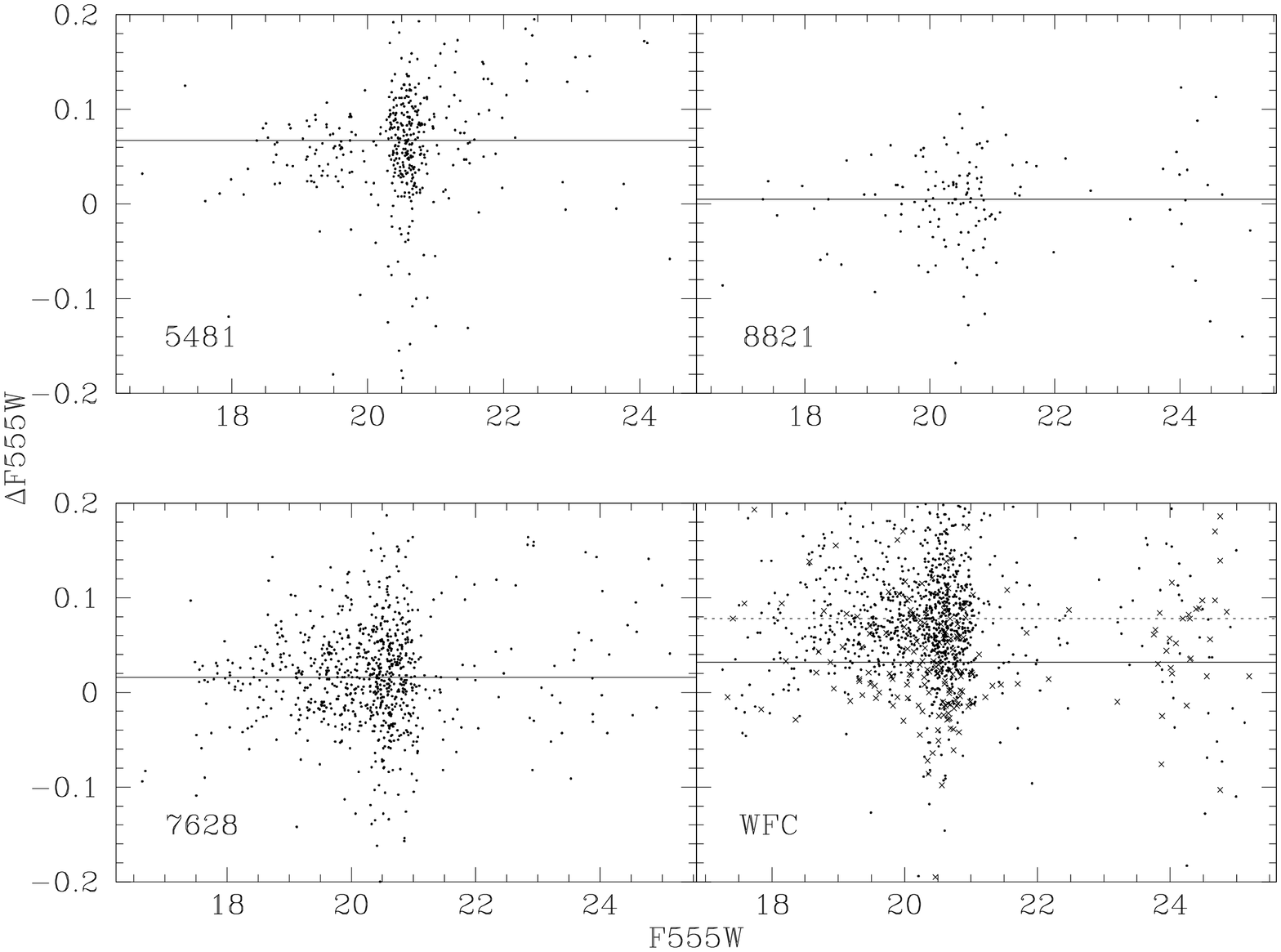}
 \caption{Residuals between F555W instrumental magnitudes
    in the sense of (7630 field - other). Lines show the median residuals. For
    the ACS WFC photometry, the {\it solid line} and $\times$ symbols are from
    the WFC2 chip (image extension 1), and {\it dotted line} and $\bullet$
    symbols relate to the WFC1 chip (image extension 2). \label{rsdv}}
\end{center}
\end{figure*}

\begin{figure*}[p]
  \includegraphics[scale=0.8]{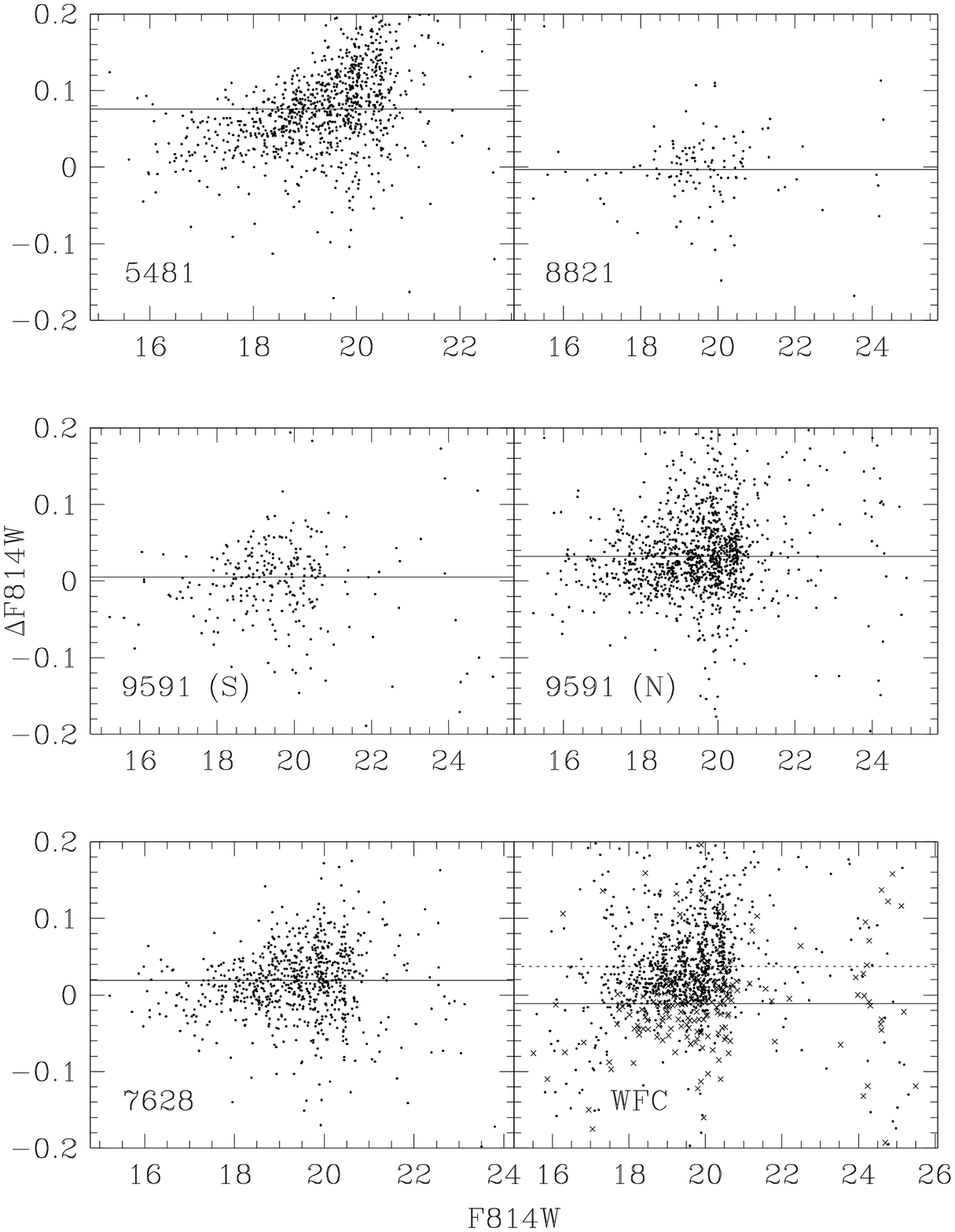}
 \caption{Residuals between F814W instrumental magnitudes
    in the sense of (7630 field - other). Lines show the median residuals. For
    the ACS WFC photometry, the {\it solid line} and $\times$ symbols are from
    the WFC2 chip (image extension 1), and {\it dotted line} and $\bullet$
    symbols relate to the WFC1 chip (image extension 2). \label{rsdi}}
\end{figure*}

\begin{figure*}
\begin{center}
  \includegraphics[scale=0.5,angle=-90]{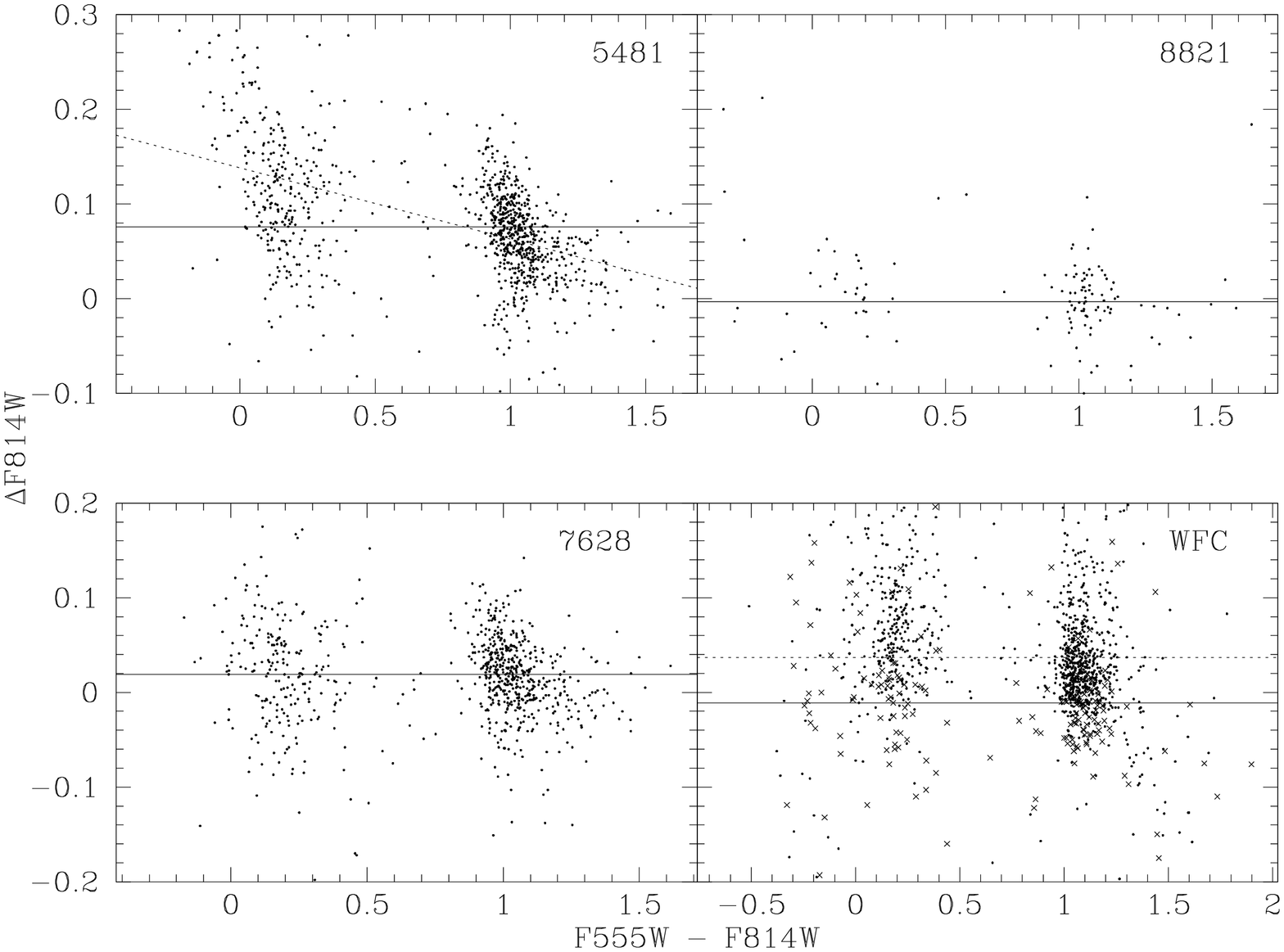}
 \caption{Residuals between F814W instrumental
    magnitudes in the sense of (7630 field - other). Points and lines have the
    same meaning as in Fig. \ref{rsdi} except that the fitted color-dependent
    trend in the 5481 data is shown.
    \label{rsdicol}}
\end{center}
\end{figure*}

In the F555W filter, we found no signs of color dependence among the
observations in the different datasets, so we only corrected for zeropoint
differences. The offsets (labeled $c_0$) are provided in Table \ref{caltab},
and are in the sense of (7630 - dataset). Residuals versus F555W magnitude are
plotted in Fig. \ref{rsdv}. In the F814W filter, we found a significant color
dependence in the residuals between the 5481 and 7630 photometry. Note that
all of the WFPC2 photometry was rereduced using the same software and
photometry parameters, so this difference appears to come from instrumental
effects. Color-dependent terms are labeled $c_1$ in the table.  Residuals
versus F814W magnitude are plotted in Fig. \ref{rsdi}, and the dependence on
color is shown in Fig. \ref{rsdicol}. We note that the color dependence of the
transformation from the WFC to the WFPC2 in the F555W and F814W filters
evaluated by \citet{sir} are consistent with zero (to within 1 $\sigma$) for
all color choices.

\begin{deluxetable}{lrrrrr}
\tablewidth{0pt} 
\tablecaption{Photometric Transformation to Proposal 7630 Photometry}
\tablehead{\colhead{Prop. ID} & \multicolumn{2}{c}{F300W} &
  \colhead{F555W} & \multicolumn{2}{c}{F814W} \\
 & \colhead{$c_0$} & \colhead{$c_1$} & \colhead{$c_0$} & 
 \colhead{$c_0$} & \colhead{$c_1$}\label{caltab}}
\startdata
5481 &       &        & $-0.068$ & $-0.141$ & 0.0778 \\
7628 & 0.113 & 0.0201 & $-0.017$ & $-0.019$ & \\
8821 &       &        & $-0.005$ &   0.004  & \\
9591 (S)\tablenotemark{a} &       &        &          & $-0.005$ & \\
9591 (N)\tablenotemark{a} &       &        &          & $-0.032$ & \\
10815 (WFC1-E)\tablenotemark{b} &       &        & $-0.033$ &   0.011  & \\
10815 (WFC2-W)\tablenotemark{b} &       &        & $-0.078$ & $-0.037$ & \\
\enddata
\tablenotetext{a}{The two proposal ID 9591 fields are labeled south or north
  based on their positions, and both are shown in Fig. \ref{pos}.}
\tablenotetext{b}{The fields of the two ACS WFC chip fields in proposal ID
  10815 fields are labeled east or west
  based on their positions, and are shown in Fig. \ref{pos}.}
\end{deluxetable}

\begin{figure*}
\begin{center}
  \includegraphics[scale=0.5,angle=-90]{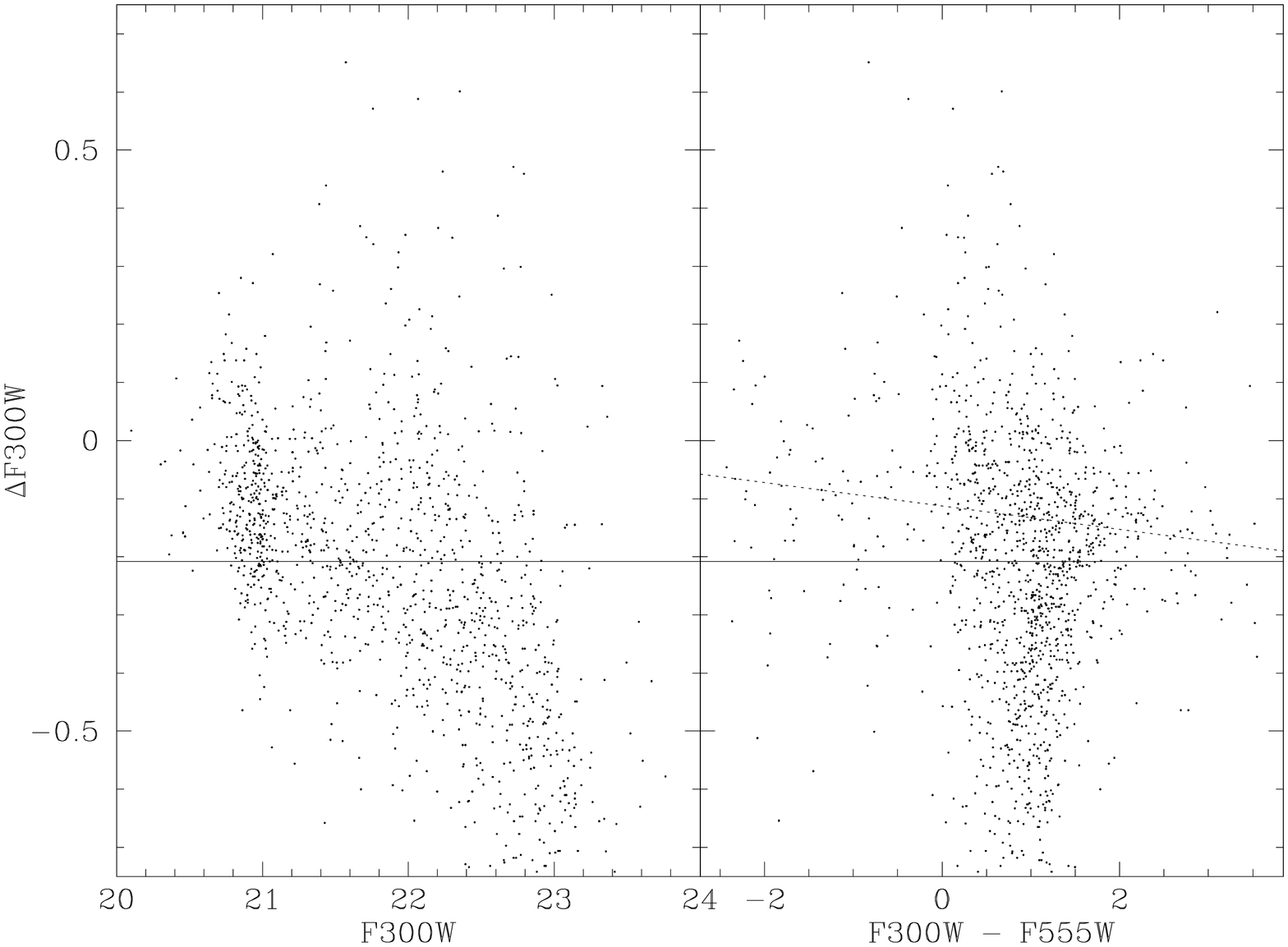}
 \caption{Residuals between F300W instrumental magnitudes 
in the sense of (7630 field - 7628 field). {\it Solid lines} show the median
residuals, and the {\it dotted line} shows the fitted trend with color.
\label{rsd300}}
\end{center}
\end{figure*}

We found a fairly large zeropoint difference and color dependence of the
photometric residuals in the F300W filter. Time-dependent zeropoint changes
are well-known in ultraviolet filters \citep{holtz}, but the behavior in F300W
has not been thoroughly mapped out because the filter is infrequently used. We
have therefore only corrected for relative differences between the two F300W
datasets, again using the proposal 7630 photometry as the standard. (Of all of
the filters considered here, the ultraviolet filters are most likely to have
significant zeropoint differences in comparison to the flight photometric
standard system.) The difference in F300W magnitudes as a function of
magnitude and color are shown in Fig. \ref{rsd300}.

\begin{figure}
  \includegraphics[scale=0.4]{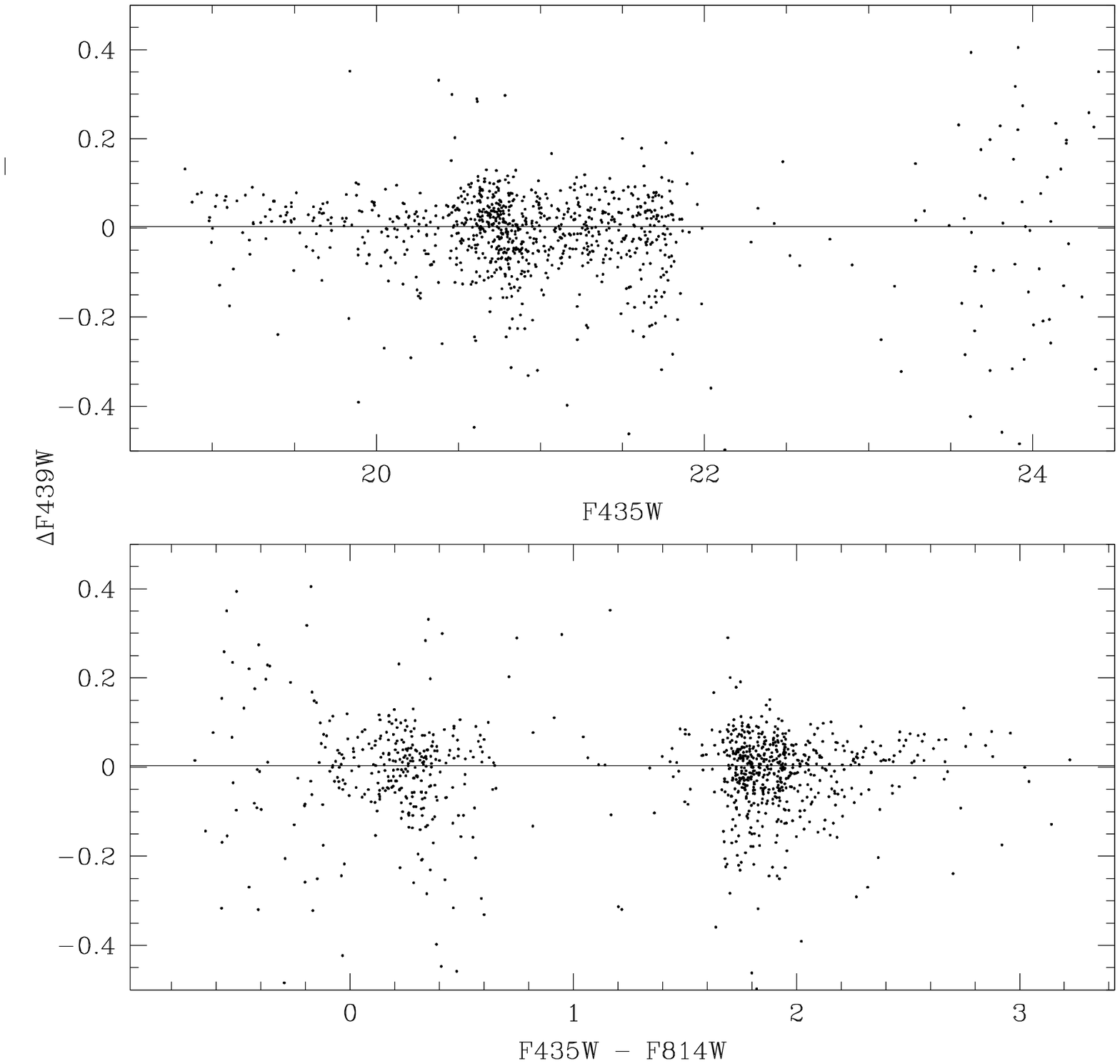}
 \caption{Residuals between WFPC2 F439W photometry
    (proposal 7628) and transformed ACS WFC F435W photometry in the sense of
    (WFPC2 - WFC). The {\it solid line} shows the median residual. Zero point
    differences between the WFC chips were less than 0.004 mag, and
    were not corrected. \label{rsdb}}
\end{figure}

The WFC F435W and WFPC2 F439W filters are also close in wavelength coverage,
although the NGC 2419 field was only observed in these filters using WFPC2 in
proposal 7628.  We compared the photometry, and found a small color term is
able to do a good job of transforming the two. The color term ($c_0 = 0.030$)
is slightly larger than the one given in Table 20 of \citet{sir}, but does a
better job at simultaneously bringing the HB and RGB populations into
consistency. Fig. \ref{rsdb} shows the comparison between the WFPC2 and
transformed WFC data.

We have generally left our photometry in the VEGAMAG flight system. When
transformations to $UBVRI$ bandpasses are needed for comparisons to external
photometry for other clusters, we have used transformation equations from
\citet{sir} for ACS WFC images, and the most recent transformations maintained
by A.  Dolphin\footnote{\tt http://purcell.as.arizona.edu/wfpc2\_calib/} for
WFPC2 images. The flight system photometry is presented in Table
\ref{phottab}.

\section{Analysis}

\subsection{The Horizontal Branch}\label{HB}

The HB morphology has been recently described (based solely on ACS photometry)
in \citet{dale}, but the focus of that study was on the blue stragglers. Here
we attempt to fully characterize the evolved populations, and in that vein we
have identified 994 HB stars in the union of the HST fields. Our $F435W$ and
$F814W$ magnitudes from the ACS data are
approximately equivalent to the $B$ and $I$ magnitudes from \citeauthor{dale}.
For most of the discussion below, the two pairs can be used almost
interchangeably (color transformation coefficients are small; \citealt{sir}).
There are a number of interesting features of the CMDs deserving of comment.

\begin{figure*}
  \includegraphics[scale=0.8]{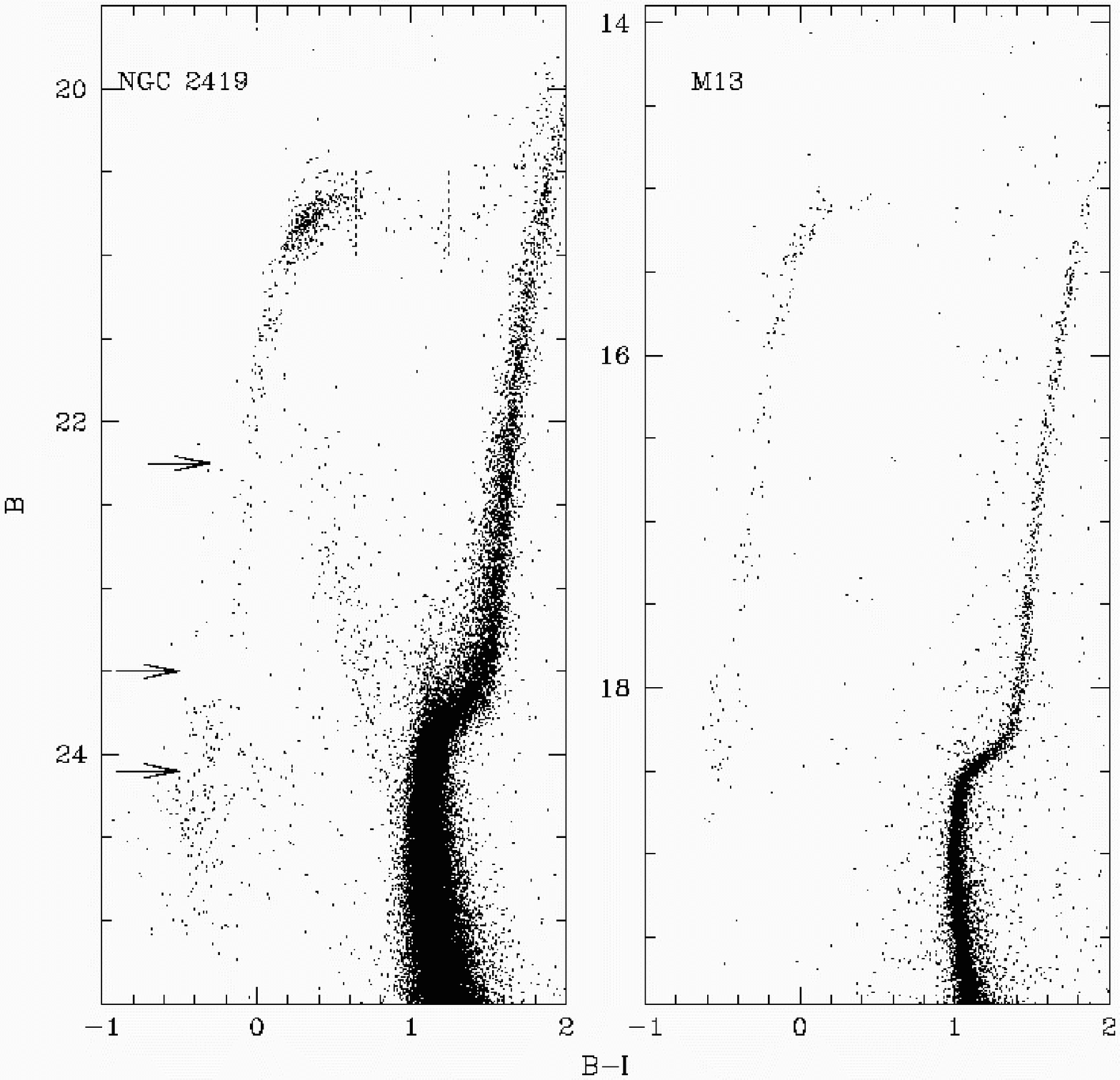}
 \caption{Comparison of the color-magnitude diagrams
    for NGC 2419 and M13 (Gordon et al., in preparation).  NGC 2419 data comes
    from ACS WFC images, and F435W and F814W magnitudes were transformed to
    Johnson $B$ and Cousins $I$ using
    equations from \citet{sir}. Only the best measured stars
    ($\vert$SHARP$\vert < 0.5$, CHI $< 5$, and crowding effects totalling less
    than 0.3 mag) have been plotted for NGC 2419. The data for M13 has been
    windowed to place the blue HB (near the instability strip) level with the
    corresponding portion of NGC 2419's HB. Arrows show the positions of
    possible gaps on the HB, and dotted lines show M5's instability strip
    \citep{sb}, shifted for the higher reddening of NGC 2419 ($\Delta(B-I) =
    0.18$) .\label{m13comp}}
\end{figure*}

\begin{figure*}
  \includegraphics[scale=0.5,angle=-90]{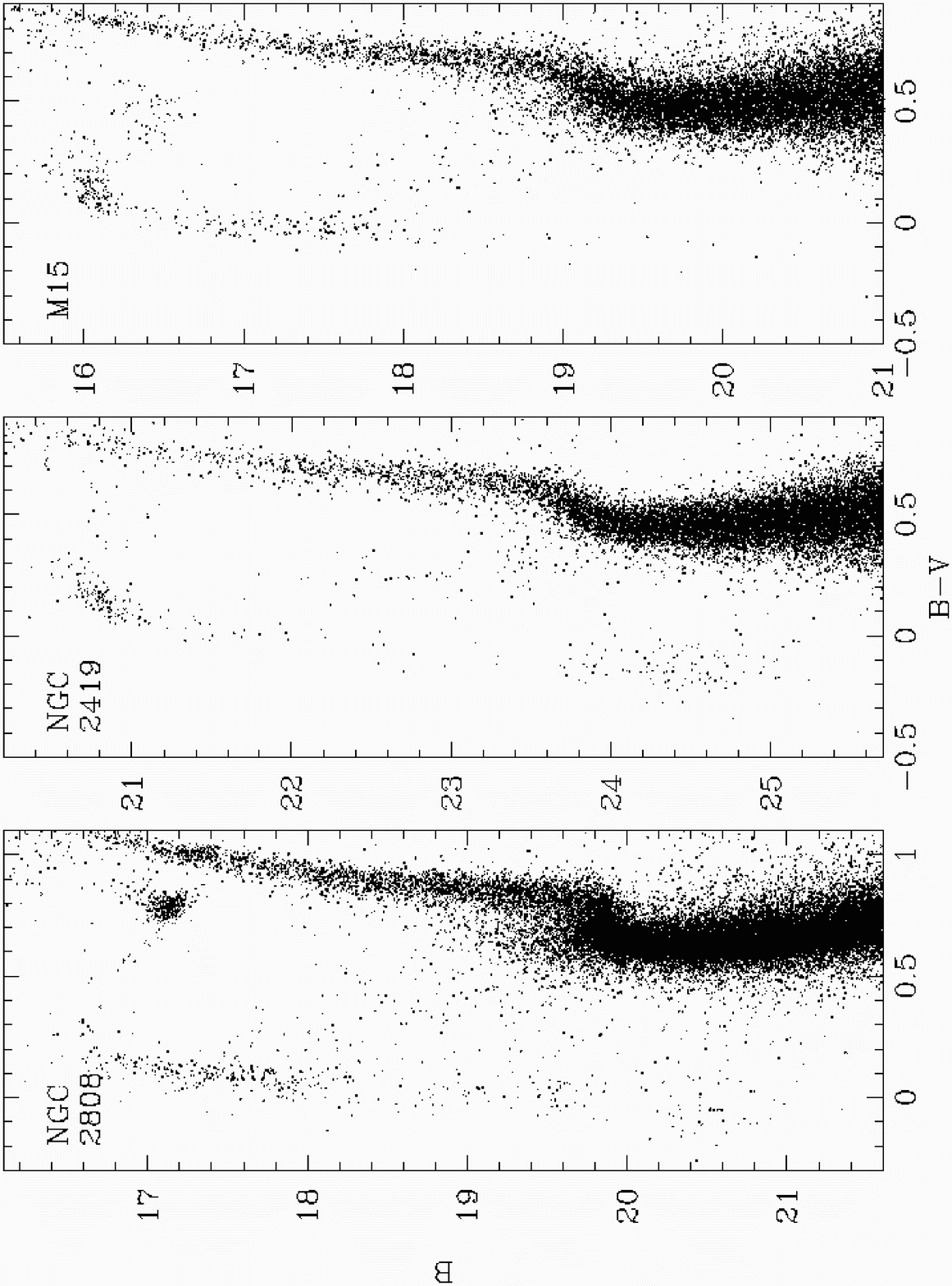}
 \caption{Comparison of the color-magnitude diagrams
    for NGC 2808, NGC 2419 and M15. NGC 2419 data comes from ACS WFC images,
    and was transformed to Johnson $BV$ using equations from \citet{sir}. The
    same quality cuts were applied to the NGC 2419 photometry as in Fig.
    \ref{m13comp}, with an additional restriction on distance from the cluster
    center ($r < 40\arcsec$). Photometry for NGC 2808 (proposal ID 6095; P.I.
    Djorgovski) and M15 (proposal ID 5324, P.I.  Yanny) comes from WFPC2 and
    was transformed using equations within HSTPhot (December 20, 2004 update;
    see {\tt http://purcell.as.arizona.edu/wfpc2\_calib/2004\_12\_20.html}).
    The data for NGC 2808 and M15 has been windowed to place the blue HB (near
    the instability strip) level with the corresponding portion of NGC 2419's
    HB.\label{btcomp}}
\end{figure*}

The HB has an extremely large total extent in effective temperature, although
the red end of the HB and the RR Lyrae instability strip are sparsely
populated, as is theoretically expected for the cluster's metallicity. There
are at least four clear breaks in the distribution of stars that can be seen
in Fig. \ref{m13comp}:
\begin{enumerate}
\item $B-I = 0.7$: This break approximately corresponds to the
  blue end of the instability strip, but there is a clear drop off in the
  number of stars redder than this point. Although we have been able to
  identify RR Lyrae variable stars from the datasets, we generally do not have
  good enough coverage of the light curve to derive precise average
  magnitudes.
\item $B \approx 21$): while not a gap in the distribution,
  this marks the blue edge of the most heavily populated part of the HB. The
  HB is more sparsely populated out to the next fainter break.
\item $B \approx I \approx 22.25$: This is the brighter of the two
  clear gaps in the HB distribution. We will refer to the stars with $22.25 <
  B < 23.5$ as the {\it inter-gap population}.
\item $B \approx 23.5$: The fainter of the two gaps marks
  the bright edge of the collection of extreme HB (EHB) and blue hook (BHk)
  stars identified by \citeauthor{dale}. Those authors tentatively identified
  this as a gap.
\end{enumerate}
\citeauthor{dale} compared the HB distribution with the more massive cluster
$\omega$ Cen, and found similarities in the overall distribution --- notably
comparable percentages of stars in the EHB/BHk group. The comparison with
$(B,B-I)$ data for M13 in Fig. \ref{m13comp} clearly shows that NGC 2419's HB
extends approximately 1 mag fainter than the greater part of the EHB clump for
M13 (although a handful of stars have been identified fainter than the clump).
A comparison with two other clusters with long blue HB tails (NGC 2808 and
M15) is shown in Fig. \ref{btcomp}. In each case, the number of HB stars
appears to sharply drop at approximately the same magnitude (relative to the
edge of the RR Lyrae instability strip): $B \sim 21$ in NGC 2808, and $B \sim
25.1$ in NGC 2419.  M15 is a cluster of similar [Fe/H] and comparable but
lower mass that converts only a has a tiny proportion of its stars into EHB
stars.  NGC 2808 has a mass that is considerably closer to that of NGC 2419,
but it has a much higher [Fe/H]. It produces more EHB stars than M15, but the
EHB stars are still a small fraction of the total HB stars.

\begin{figure}
  \includegraphics[scale=0.4]{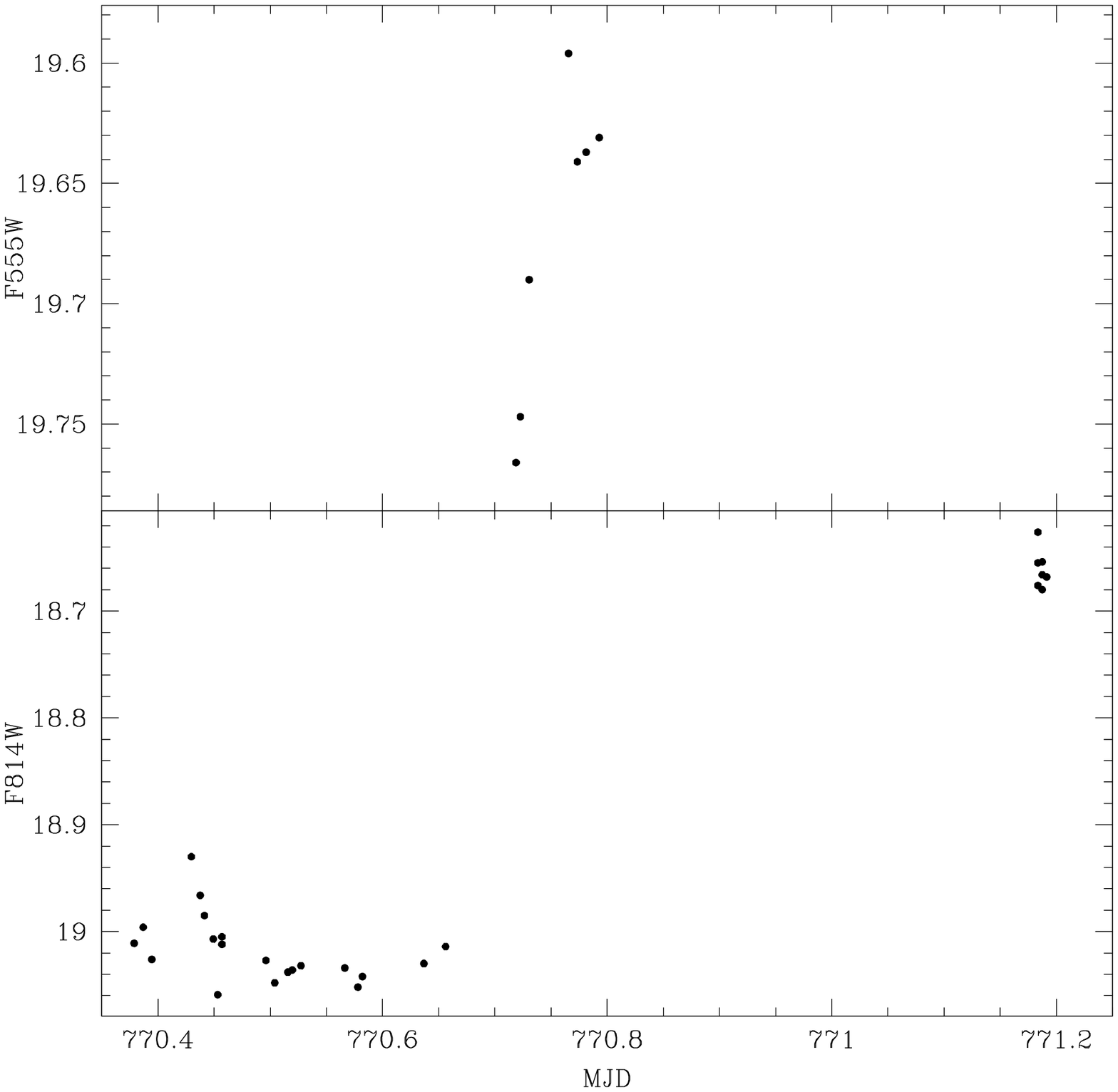}
 \caption{The light curve for the bright RR Lyrae variable
    V2 from {\it HST} photometry from proposal ID 7630. The time coordinate is
    modified Julian date (heliocentric Julian date minus 2450000).\label{v2}}
\end{figure}

The majority of the cluster's HB stars fall in a compact group in the CMD
($20.6 < B < 21$, $0.2 < B-I < 0.6$; see Figure \ref{m13comp}), with less dense
extensions to the red and blue. There is a fair amount of evidence to suggest
that all or nearly all of the stars redward of the main peak have evolved to
their present colors from the main peak. The presence of AGB stars (especially
ones in a ``clump'' at the base of the AGB) implies redward evolution after
the HB phase, and at least some models (see the next paragraph) predict stars
should spend a substantial time redward of their starting point on the HB.
Probably the strongest evidence comes from the comparison of the
period-amplitude diagram for NGC 2419 RR Lyrae stars with that of M3
\citep{rip}. M3 is a cluster which has strong populations of stars both redder
and bluer than the instability strip, and so it is legitimate to expect from
theoretical evolutionary tracks that there are relatively unevolved RR Lyrae
stars as well as stars that have evolved into the instability strip from the
blue HB. The lower average densities of the evolved stars are expected to
result in larger periods, and this is supported by the appearance of two
sequences in the period-amplitude diagram \citep{m3}.  NGC 2419 RR Lyraes fall
nearly exclusively among the ``evolved'' population of M3 stars (see Fig.  3
of \citealt{rip}).  In this context, star V2 appears to be an unusually bright
evolved RR Lyrae star. The CMD position identifies it as a ``supra-HB'' star
(see \S \ref{uvsec}). We have identified V2 from the finding chart published
by \citet{prrrlyr}, and from our own detection of its variability in the {\it
  HST} data. Our photometry implies that it was observed near minimum light in
the F814W filter (see Fig.  \ref{v2}). The only published light curve
\citep{prrrlyr} implies it has a large amplitude (about 1.3 $B$ magnitudes)
and period, although their photometry indicates that it is comparable to other
known RR Lyrae stars oscillating in the fundamental mode.  At the same time,
their photometry implies that it is also similar in brightness to other RR
Lyraes as well.  This contradiction between the \citeauthor{prrrlyr}
photometry and ours suggests that this star should be studied in more depth.

HB sequences (e.g. \citealt{dsep}) imply that zero-age horizontal branch
(ZAHB) stars for NGC 2419's low metallicity and primordial helium abundance
always fall bluer than the instability strip, even for stars with no mass
loss.
Synthetic HB models also predict a significant number of variable and red HB
stars for a zero-age population that begins at the position of the main peak
(see Fig. \ref{hbtracks}).  Blueward evolution in HB stars tends to occur in
stars with lower initial $L_{He} / L$ \citep{ir}, where $L_{He}$ is the
luminosity released by helium fusion.  This means that the hydrogen-fusion
shell rapidly adds mass to the helium core of the star. This can result from
higher helium abundances or metallicities (both affecting the structure of the
envelope, and producing higher shell fusion temperatures), but in general it
requires a substantial hydrogen envelope in order to maintain the
conditions needed for a strong hydrogen-fusion shell \citep{swei}. When the
hydrogen shell is weakened (by low envelope mass or low envelope helium
abundance, for example), the color evolution is driven by the depletion of
helium in the core as the star becomes more giant-like. So the evolution of HB
stars can bear on questions of helium abundance, at least tangentially. The
relatively low dispersion in magnitude among the red HB stars is consistent
with the idea that the stars in the main peak have low (approximately
primordial) helium abundance. An increase in $Y$ of as little as 0.05 produces
an initial blue loop in the star's evolution, and the subsequent evolution
toward the giant branch takes place at a significantly greater luminosity.
(See Figure 9 of \citealt{busso} for a comparison of much more metal-rich tracks
with different helium abundances.)

The more important issue is what can be learned about the population of the
second peak on the EHB, as it is the bluer population of HB stars that is
identified with helium-enriched gas when multiple populations are present. An
understanding of this group is complicated by poorer sensitivity to
temperature of the colors available here, and the fact that some of these
stars appears to exist beyond the end of the canonical HB, as seen in Fig.
\ref{m13comp}.  Because these stars are at or beyond the end of the HB, they
are called extreme HB or blue hook stars (we will refer to them as the EHB/BHk
group).  There are major uncertainties in models of these stars: for example,
it is still unclear exactly how the stars are formed and whether the formation
mechanism has a strong effect on the position in the CMD, but the combination
of flash nucleosynthesis and convective mixing makes modeling difficult
\citep{brown}.  Regardless, in the analysis below, we first attempt to
identify as many of the stars associated with this group as possible, and then
examine the evolution paths they take.  Because the evolution is seen most
clearly in some of the ultraviolet datasets, we discuss it in \S \ref{uvsec}.

\begin{figure}
  \includegraphics[scale=0.3]{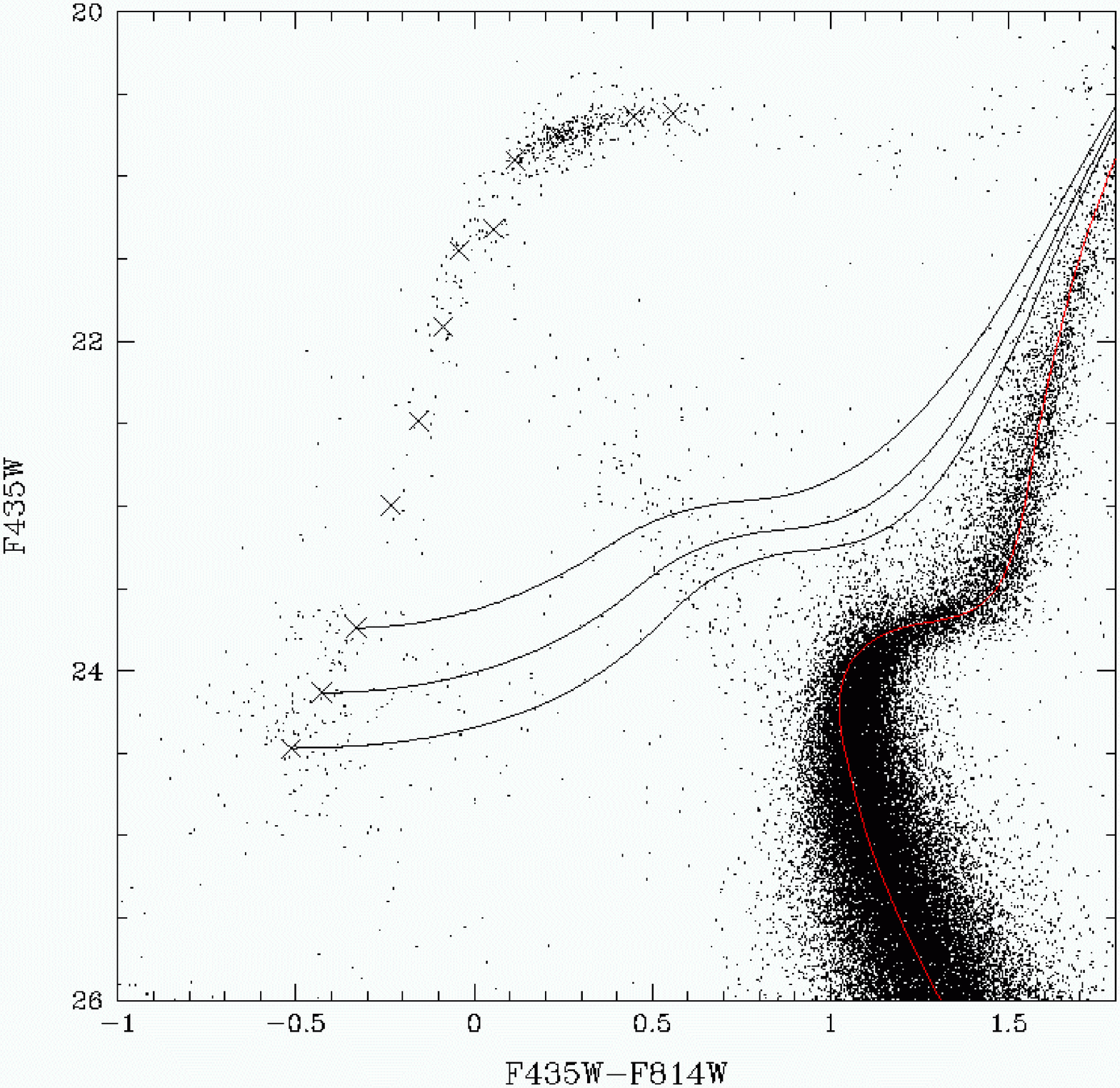}
 \caption{The effects of blending on the observed
   properties of EHB/BHk stars in the ACS WFC data.  A DSEP isochrone
   \citep{dsep} with age 13 Gyr, [Fe/H]$=-2.1$, and empirical color
   transformations was fit to the main sequence and lower giant branch ({\it
     red line}), and the curves show blends of the EHB star with a star whose
   properties were drawn from the fitted isochrone.\label{ehbblend}}
\end{figure}

In the ACS photometry (see Fig. \ref{hbr60}), we consistently see a tail of
stars stretching from the EHB/BHk group toward the faint end of the blue
straggler group and main sequence turnoff ($23.6 < F435W < 25$, $-0.3 <
F435W - F814W < 0.5$). These are probably stars that are
optical blends of an EHB/BHk star with a main sequence star. \citet{castel}
saw similar features in their optical CMD of NGC 2808. The features were not
as strong there, probably because NGC 2808 has a much smaller EHB population
producing such objects in the CMD. Fig. \ref{ehbblend} shows loci of blends of
three EHB/BHk stars (marked by the faintest 3 $\times$ symbols) with main
sequence and giant branch stars [represented by a \citealt{dsep} (hereafter,
DSEP) isochrone having age of 13 Gyr and [Fe/H]$ = -2.1$ that was fitted to
the stellar locus]. Blends with turnoff stars fall near the blue straggler
region near the kinks in the curves. While such blends could masquerade as
blue stragglers in optical CMDs, they can be clearly identified in ultraviolet
CMDs due to the much higher surface temperatures of the HB star.

\begin{figure}
  \includegraphics[scale=0.4]{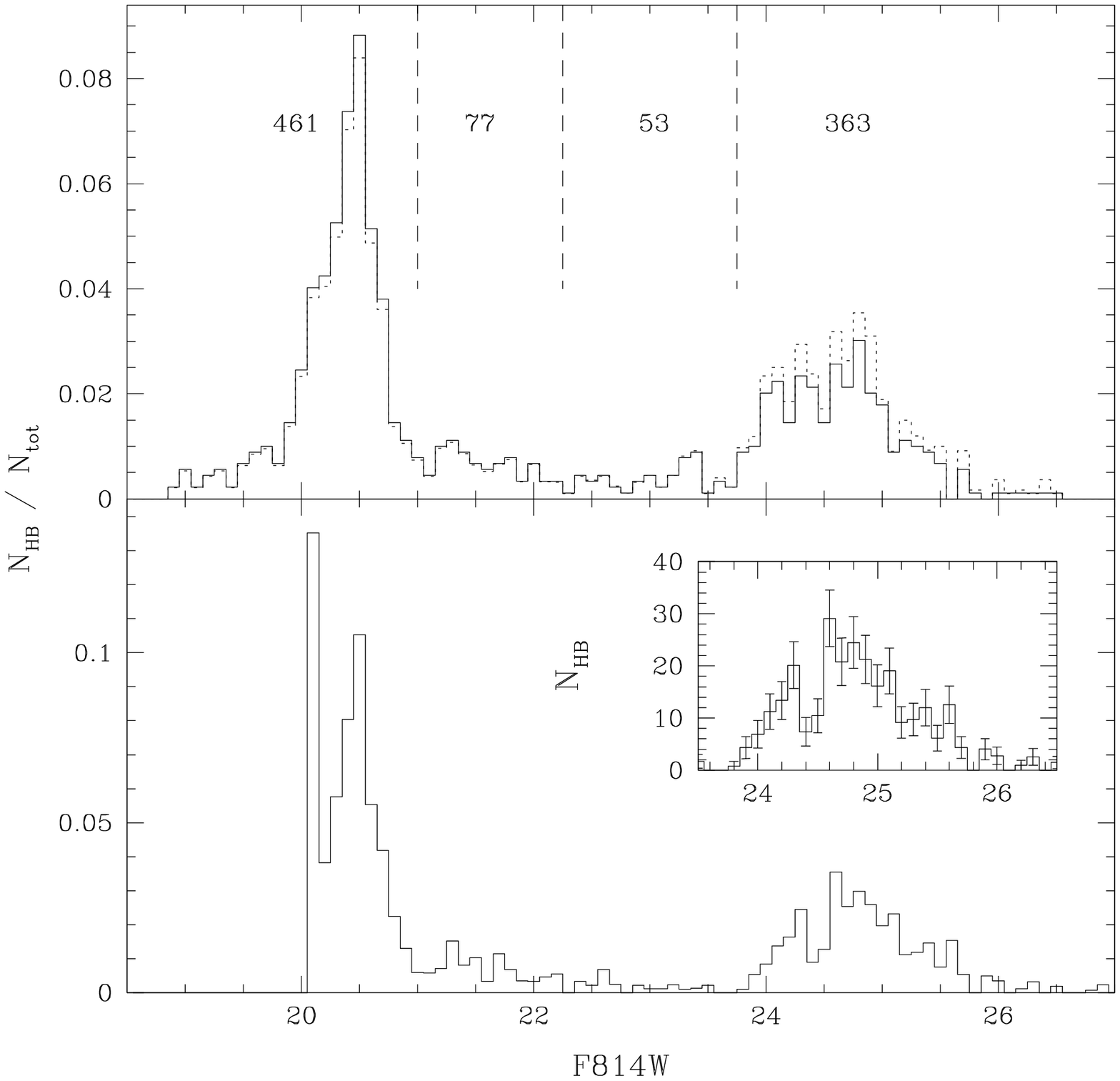}
 \caption{{\it Top panel:} Detected {\it (solid line)}
    and incompleteness-corrected {\it (dotted line)} HB star distributions for
    stars in the ACS fields. The numbers refer to incompleteness-corrected
    numbers of stars in each of the magnitude ranges separated by dashed
    lines. {\it Bottom panel:} HB star distribution corrected for lifetime
    differences, and including deprojections of evolved stars from the main
    peak ($F814W = 20.1$) and probable EHB/MS blends.  {\it Inset:} Number
    of EHB stars, corrected for incompleteness and blending, and including
    Poisson uncertainties on each bin.
    \label{ihist}}
\end{figure}

We use the results of the artificial star tests (\S \ref{artstar}) to correct
the distribution of detected HB stars for incompleteness.  We took the total
sample of identified HB stars and only kept stars that were detected in either
the ACS WFC or HRC photometry and passed the same quality cuts as the
artificial stars. This sample was made up of 856 stars, and with
incompleteness corrections, the total sample expected in the field was 954.
The top panel of Fig. \ref{ihist} shows the detected and
incompleteness-corrected HB distributions for the ACS fields. Based on this,
the EHB population makes up more than 38\% of the total HB population.

\subsubsection{Accounting for HB Star Lifetimes}

Theoretical models uniformly predict that
stars at the blue end of the HB have 20-30\% longer HB lifetimes than stars
near the instability strip, so we need to correct for this effect to determine
how frequently EHB stars are being produced compared to the rest of the HB
population. Because the HB population is used in determining the initial
helium abundance, corrections for differences in HB star lifetimes can make
the measurements more accurate.
Below we describe a method for correcting for variations in HB lifetimes. As
much as possible, we wish to produce a correction that i) is independent of
chemical composition so that composition variations or imperfectly known
compositions do not introduce systematic errors, and ii) is based on
observational quantities.



\begin{figure}
\epsscale{.90}
\includegraphics[scale=0.4]{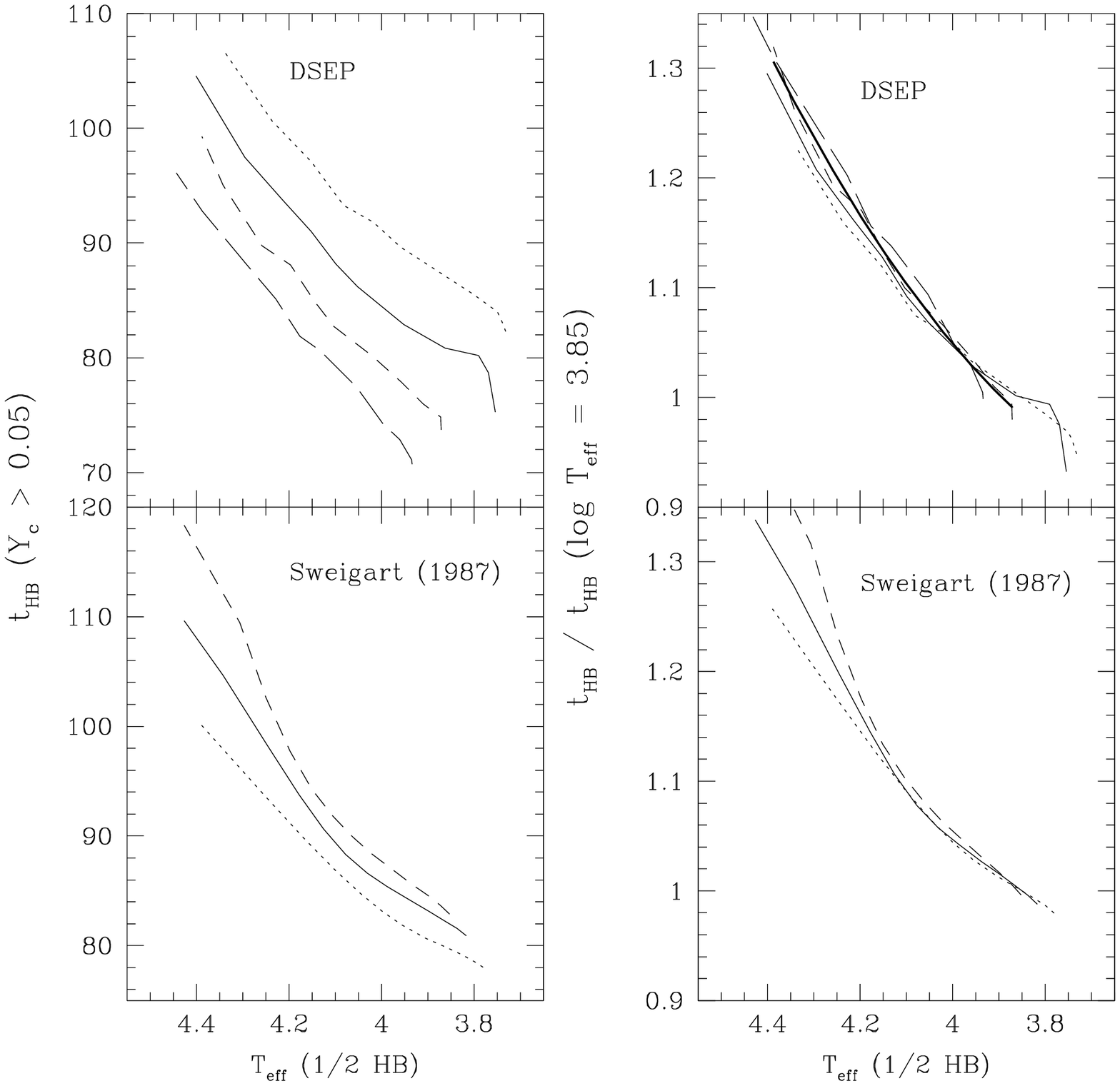}
\caption{HB lifetimes as a function of $T_{eff}$ from DSEP \citep{dsep} models
  for (from top to bottom) [Fe/H] $= -1.0, -1.5$, $-2.0$, and $-2.5$
  ([$\alpha$/Fe] $= +0.2$), and \citet{swei} models $Y = 0.30, 0.25$, and 0.20
  ($Z = 0.0001$).{\it Left panels:} lifetimes in Myr. {\it Right panels:}
  lifetimes normalized to the stars reaching the zero-age HB at $\log T_{eff}
  = 3.85$.  (For the DSEP model with [Fe/H] $\le -2.0$, the most massive
  models do not reach that temperature, and were normalized at the lowest
  effective temperature present in the model.) In the top right panel, the
  bold line shows the polynomial fit to the [Fe/H]$ = -2.0$ models that was
  used in making lifetime corrections.
\label{hblife}}
\end{figure}

For stars with the same initial compositions, we find that the lifetimes vary
in similar ways as a function of stellar mass and effective temperature,
although the absolute value of the lifetime does depend on composition. (In
the following we have defined the end of the HB phase as the point when the
central helium abundance $Y_c = 0.05$; see the lefthand panels of Fig.
\ref{hblife}. Because HB stars evolve somewhat in temperature during this
phase, we have used the temperature of the star when it is halfway through its
HB lifetime as a representative $T_{\rm eff}$. We can imagine more complex
algorithms for determining this temperature, but even the use of the $T_{\rm
  eff}$ at zero-age produced minor effects on the results.) Based on this
behavior, we defined a weighting factor for each star in the HB sample:
\[w_i = \frac{t_{HB}(\log T_{eff} = 3.85)}{t_{HB}(\log T_{eff})} \]
The largest portion of the variation in HB lifetimes occurs on the blue tail
for $T_{eff} \ga 10^4$ K in stars with low-mass hydrogen envelopes. By
choosing $\log T_{eff} = 3.85$ (near the blue edge of the instability
strip) as the normalization point, we can correct the blue HB lifetimes
to values representative of variable and red HB
stars, which are common in globular clusters.
The variation of the weighting factors with $T_{eff}$ is shown in the
righthand panels of Fig. \ref{hblife}. There is very little variation in the
weighting factors with variations in [Fe/H], so that residual uncertainties
are at the level of a few percent. Further, the weightings {\it only} remove
the effects of position on the HB, and do not reference the absolute value of
the HB lifetime, which does depend on composition and physics inputs to the
stellar evolution codes (e.g., the $^{12}$C($\alpha,\gamma)^{16}$O reaction
rate; \citealt{cass03}). 

The weighting factors also vary little with $Y$ for $\log T_{\rm eff} < 4.2$,
although the agreement gets worse at higher $T_{\rm eff}$. Higher helium
abundance produces HB stars whose lifetimes increase more rapidly with
decreasing $T_{\rm eff}$ than stars with lower helium. So for clusters with
the majority of their HB stars having $T_{\rm eff} \la 16000$ K, the weighting
factors are almost entirely independent of composition, and corrected HB star
totals can be used to gauge helium abundance directly using the $R$ ratio. For
clusters like NGC 2419 that instead have a larger fraction of their stars at
higher $T_{\rm eff}$, this is not as straightforward, but the $R$ value can
still be used as a test of whether clusters stars have enhanced helium. In
fact, if weighting factors having a canonical value of $Y \approx 0.25$ are
used, then helium-enriched stars on the blue tail will be undercorrected for
their long lifetimes. This has the effect of leaving the corrected number of
HB stars and the $R$ ratio too large, and implying a high $Y$.

The weighting has the most utility when dealing with stellar populations that
have the same composition. Although NGC 2419 has not shown any evidence of
large star-to-star composition ([Fe/H], $Y$) differences in its CMD (via
large dispersion in its evolutionary sequences or multiple populations), other
massive clusters show clear signs of internal chemical enrichment. Further,
very blue HBs may result from helium enrichment.  In order to look for signs
of helium enrichment in NGC 2419 using the population ratio $R$, we need to
correct for the effect of its blue HB first (in case it has a different cause,
such as mass loss). Helium-enriched stars have longer lifetimes than
unenriched stars at a given $T_{\rm eff}$ --- at the blue end of the HB the
difference is almost 10\% for $\Delta Y = 0.05$.  Detection of a
helium-enriched subpopulation of HB stars will still depend on the relative
numbers of enriched stars, but the correction will provide a more accurate
idea of the size of such a population. Some authors have previously attempted
to correct for the lifetime effect. \citet{cass03} made corrections to $R$ for
clusters with blue HB tails by using an HB evolutionary timescale appropriate
to the mean HB mass. For clusters with bimodal HB distributions, this can lead
to significant error because the lifetime is not a linear function of mass.
\citet{casomega} attempted to improve their analysis of $\omega$ Cen by
applying corrections to the HB population using ``typical'' absolute lifetimes
for stars in three well-populated parts of the HB. We believe our method is a
significant improvement in accuracy.

\begin{figure}
\includegraphics[scale=0.4]{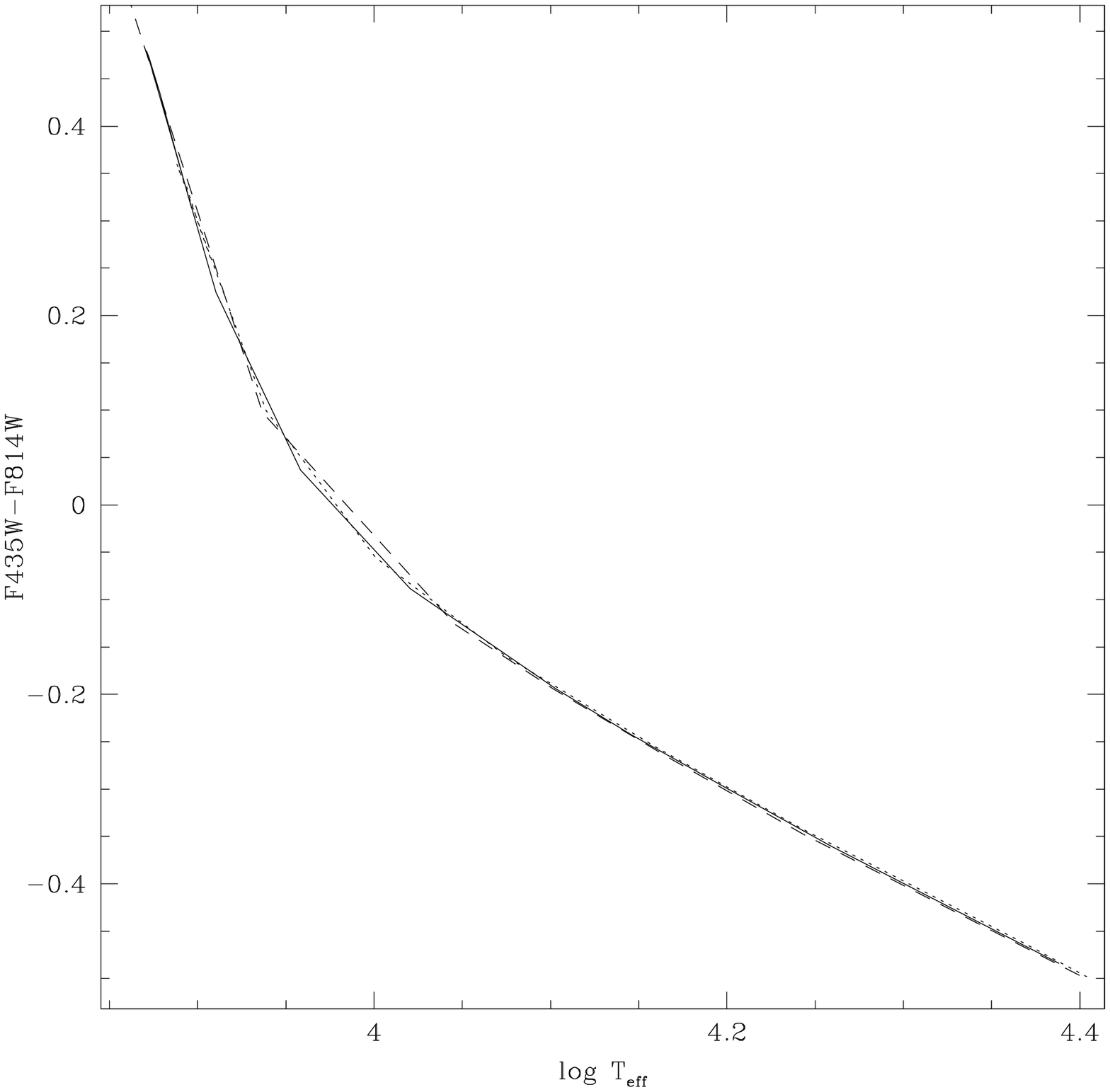}
\caption{ACS WFC color-$T_{\rm eff}$ conversion derived from DSEP \citep{dsep}
  models for [Fe/H] $= -2.0$ and [$\alpha$/Fe] $= +0.2$. The lines correspond
    to the zero-age HB ({\it dotted line}), the HB termination (core helium
    $Y_c = 0.05$; {\it dashed line}), and halfway in time between the two.
  No reddening has been applied to the color.
\label{colteff}}
\end{figure}

Although color-$T_{eff}$ relationships remain imperfect, it is much more
reliable to derive $T_{eff}$ from photometry than it is to derive stellar mass
(which is the only other plausible choice for a physically-based coordinate
describing position along the HB). To do this, we used DSEP HB models to
determine an theoretical ``average'' HB line: specifically the locus of points
halfway through the HB lives (ending at $Y_c = 0.05$) of stars of different
masses. The color-$T_{\rm eff}$ transformation we used is shown in Fig.
\ref{colteff}. This average line was fit to the stellar photometry and used to
derive the offset between the theoretical and observed colors. For this
cluster, we used the $F435W-F814W$ color (giving a color shift of 0.13) due to
its good temperature sensitivity across the color range. Note that the
evolutionary stage does not have a significant effect as long as the color and
$T_{\rm eff}$ are determined consistently --- the other atmospheric variable
affecting the colors (surface gravity) does not change greatly during the HB
phase.

\begin{figure*}
  \includegraphics[scale=0.7,angle=-90]{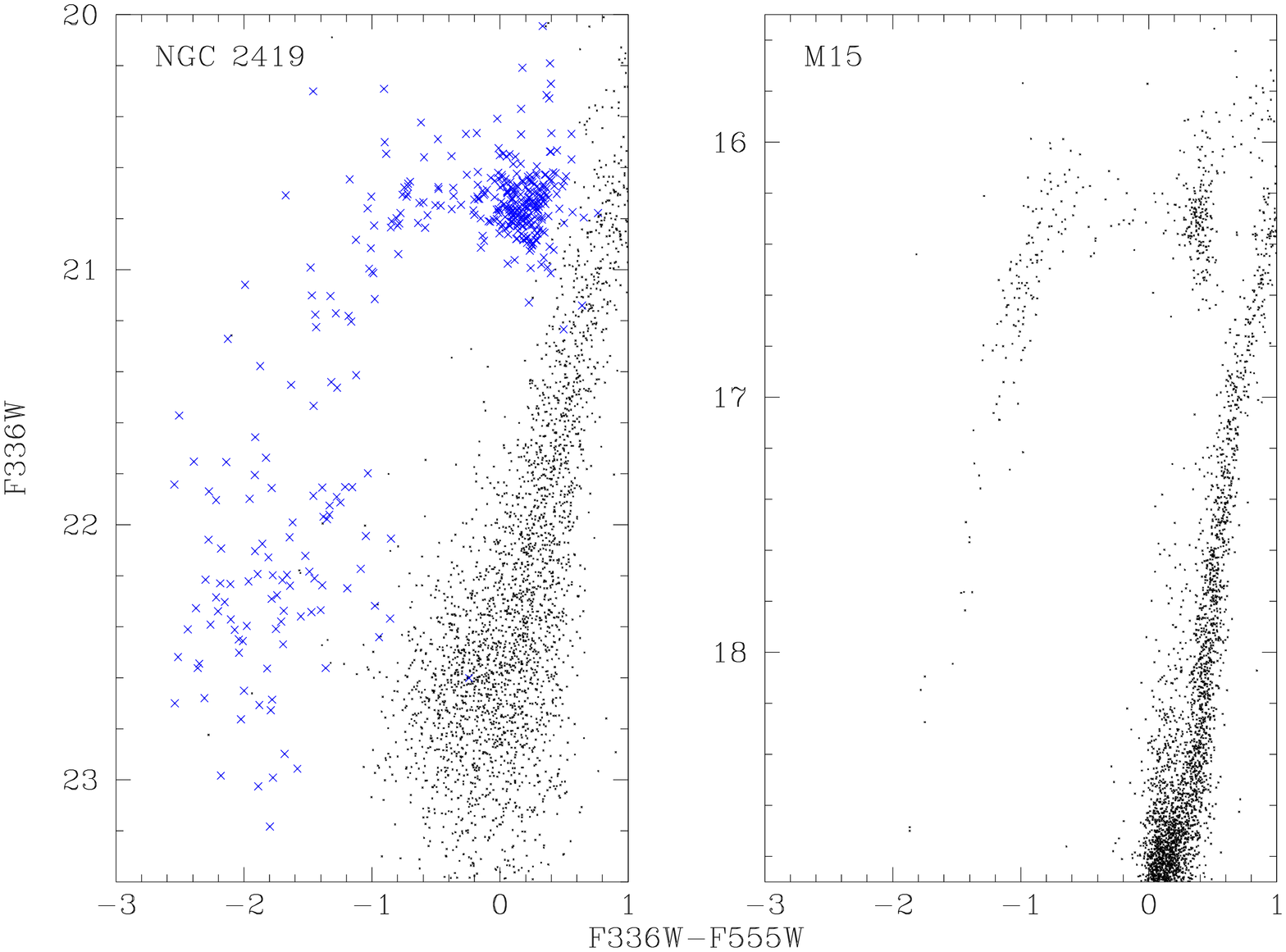}
 \caption{Comparison of the color-magnitude diagrams
    for NGC 2419 and M15 using the WFPC2 camera and F336W filter. The data
    for M15 has been windowed to place the blue HB (near the
    instability strip) level with the corresponding portion of NGC 2419's
    HB. Blue crosses are stars that were identified as HB stars via the
    methods described in \S \ref{selstar}.\label{m15f336}}
\end{figure*}

An additional benefit of using photometry to define the temperature scale is
that there are features (such as the edges of the instability strip) that can
be used as reference points. For example, the Grundahl $u$-jump has been found
in many clusters with blue HB stars with $T_{eff} \ga 11500$ K \citep{ujump}.
The $u$ jump is difficult to find in NGC 2419 because the HB is not heavily
populated near its expected position, but in M15 it can seen at
$F336W-F555W = -0.6$ (see Fig. \ref{m15f336}). For NGC 2419, the jump
appears to be at $F336W-F555W = -0.65$, which corresponds to $F439W
\simeq 21.35$ in the WFC photometry, and a small color shift can be discerned
at this position in the optical CMD. This is within the modestly populated
blue wing of the main peak in the HB distribution. Our conversion from
$F435W-F814W$ independently produced $T_{\rm eff} \approx 10950$ K,
gratifyingly close to 11500 K \citep{ujump}. ($T_{\rm eff} = 11500$ K was
predicted to be about 0.05 mag fainter.) In addition, the normal HB also seems
to have a reasonably well-defined termination near 29000 K \citep{brown08}. As
seen in Fig. \ref{m13comp}, the end of M13's HB (which appears to correspond
to the end of its hot HB) translates to the middle of the EHB clump in NGC
2419 ($F435W \approx 24$). Our color-temperature conversion again produces
temperatures that approximately ($\sim 28500 - 29000$ K) match the theoretical
value.  A temperature of 35000 K (corresponding to the transition in
atmospheric helium abundance spectroscopically observed in NGC 2808 and
$\omega$ Cen) would fall near $F814W = 24.9$ and $F435W = 24.5$.

As expected from the analysis above, the use of different sets of models will
have a small effect on the lifetime-corrected HB star total. For our final
analysis, we fitted a second-order polynomial to the weighting factors
computed from DSEP models for [Fe/H]$= -2.0$, [$\alpha$/Fe]$=+0.2$, and $Y =
0.244$ in order to smooth out numerical noise. In addition, we assumed that
stars redder than the main peak had evolved from the peak, and so were
assigned weights from the main peak. We also needed to extrapolate the
weighting corrections in the range $4.4 \le \log T_{eff} \le 4.5$, due to a
lack of tabulated models. This is somewhat risky because a discontinuous
change in the chemical composition of the atmospheres (as there is in NGC 2808
and $\omega$ Cen) would cause a discontinuity in the temperature scale derived
from the photometry. In this case we are likely to be underestimating $T_{\rm
  eff}$ and the lifetime correction. This would produce a slight bias toward
detecting a helium enhancement, but our corrections should still remove the
lifetime effect to first order. With corrections for
both incompleteness and differences in lifetimes, our original sample of 968
HB stars becomes
$N_{HB}^{\prime} = 821\pm29$. (The quoted error bar is based on Poisson
statistics for the corrected HB number.) It is important to note that this
would correspond to the number of HB stars relative to bright RGB stars in our
sample if i) the HB stars landed on the ZAHB near the blue edge of the
instability strip, and ii) the initial chemical composition of all stars was
the same. For this reason, it is appropriate to use this in the helium
indicator $R$ (see \S \ref{R}).

If the stars were formed with the same composition, then we can use the
lifetime-corrected histogram to derive
the production rate of EHB stars relative to other HB stars: 33\%.  However,
increased metal and helium abundances both increase the HB lifetimes, so any
chemical enrichment among some fraction of the cluster's stars would increase
$N_{HB}^{\prime}$ above the value for a chemically uniform population.  So the
fractional production rate of EHB stars in NGC 2419 is at most 33\%.  Fairly
large changes in composition are theoretically predicted to produce fairly
small changes in HB lifetime though: from DSEP models, a 0.5 dex change in
[Fe/H] changes the lifetime by about 5\%, and from \citet{swei} models a 0.05
change in $Y$ produces a 7\% change among stars at the hot end of the HB. Only
extreme chemical enrichment among the EHB stars is likely to mean that fewer
than 25\% of HB stars being produced are being made EHB stars.

We also note that if EHB stars are not produced at a constant rate through a
cluster's lifetime that an EHB population would take some time to
equilibrate because production rates (relative to other HB stars) would
not initially equal ``destruction'' rates (stars evolving off the EHB). EHB
populations could be expected to linger an extra 30 Myr or so after the
production mechanism (whatever it might be) is shut off.

\subsubsection{An EHB Gap?}

\citet{brown} presented evidence that the faint gap in NGC 2808 ($B \approx
20$; see Fig. \ref{btcomp}) corresponds to a theoretically-expected gap
between EHB stars and blue hook stars which ignite helium while on the white
dwarf cooling curve. \citet{moe2808} find that the gap in NGC 2808 corresponds
to a spectroscopic transition from stars with helium-poor atmospheres
(brighter than the gap) to ones with helium rich atmospheres. They find that
the brighter stars have temperatures and gravities consistent with normal HB
evolution, and that helium has probably diffused out of their atmospheres. The
fainter, helium-rich stars have properties consistent with a mixed envelope
associated with a late helium flash. In addition, \citeauthor{moe2808} found
that stars in $\omega$ Cen show a transition in atmospheric helium abundance
at the same $T_{\rm eff}$ ($\sim 35000$ K), and \citet{moem15} found a faint
HB star in M15 ($B = 19.75$) with a helium rich atmosphere, and this star also
falls below the level of the gap in NGC 2808.  However, \citet{brown} also
noted that a gap similar to the one in NGC 2808 does not clearly show up in
the photometry of some of these same clusters, so we have examined this
question again here.

Blue hook candidates have been identifed in several clusters through a
comparison of CMDs.  Photometry in the F255W filter appears to give some of
the clearest evidence of BHk stars. Clusters like M13 and M80 have HBs that
terminate approximately 1.4 magnitudes below the knee (the blue HB stars with
maximum brightness in F255W), and thought to have EHB stars but no BHk stars
\citep{ferr98}.  \citet{dale08} used ($F255W, F255W-U$) CMDs to compare
NGC 6388 and M80, and show that NGC 6388 had a group of stars that were
fainter than the faintest HB stars in M80 and were separated from the rest of
the HB stars by a gap. Photometry for NGC 2419 in the F250W filter from HRC
camera (Fig. \ref{f250}) places the bright end of the EHB about 0.9 mag
fainter than the knee and the presumptive EHB gap about 1.1 mag below the
knee. Because the F250W and F255W bandpasses are not identical and the
photometry for these hot stars are sensitive to it, we need a more exact
comparison.

We used the ``$B$ knee'' as a reference point to shift the CMDs of the
different clusters, as did \citet{dale} in their comparison of NGC 2419 with
$\omega$ Cen. The knee is essentially based on surface temperature. NGC 2808's
HB is not well populated at the knee, which could produce some error, but its
red HB allows us to check that the alignment with M15 is not far off. The most
clearly visible gaps in NGC 2419's HB distribution do not match up with those
in other clusters when the HB ``knees'' are aligned, as in Figs.
\ref{m13comp} and \ref{btcomp}. The possible exception is a marginally visible
gap at $B \approx 24.1$ ($I \approx 24.5$) in NGC 2419. Its presence is masked
in the upper panel of Fig.  \ref{ihist} because of the presence of probable
EHB/MS blends. In the lower panel of that figure we have therefore deprojected
stars back to the HB fiducial line using the blending lines shown in Fig.
\ref{ehbblend}. After this procedure is executed as part of the incompleteness
correction process, we see clearer evidence of a gap.  The statistical
significance of such a feature is hard to gauge because there is no
theoretical expectation for how the stars should be distributed, but in the
inset we plot corrected star numbers with Poisson distribution uncertainties.
The bins at $F814W = 24.4$ and 24.5 appear to be several standard deviations
below the general trend defined by bins on either side.  However, experiences
with other clusters \citep{gaps} lead us to believe that there is a
significant chance that it could be produced by statistical fluctuations even
with the relatively large sample of stars we have in NGC 2419. However, NGC
2419 is not the only cluster to show a feature here.

As shown in Fig.  \ref{m13comp}, M13 has produced only a few possible HB stars
hotter than the clump at the end of its blue tail \citep{ferr97}, which
terminates at the level of the weak gap in NGC 2419. In Fig. \ref{btcomp}, M15
shows a scattering of stars in the EHB (the one spectroscopically identified
BHk candidate falls below the level of NGC 2808's gap).  $\omega$ Cen has a
large population of EHB stars with no gap, although \citet{brown} mention that
one might have been blurred by the wide metallicity dispersion from star to
star.  Still, a feature {\it can} be seen in the \citet{casomega} HB
distribution for $\omega$ Cen at $B \sim 18.5$. The distributions for $\omega$
Cen and NGC 2419 are compared side-by-side in Fig.  4 of \citet{dale}, shifted
in the same way as in Fig. \ref{btcomp}. The feature appears there not so much
as a gap, but as a strong contrast in CMD density: the fainter blue hook stars
are much more common than the slightly brighter EHB stars.  Using $V$-band and
shifting the CMDs to account for differences in distance and reddening,
\citet{rosem54} found that a gap on the extreme blue tail of M54 precisely
overlaps the one in NGC 2808. The end of the theoretical zero-age HB fitted to
NGC 2419 in Fig. \ref{hbtracks} also appears to terminate at the approximate
position of the apparent gap. All of this leads us to postulate that the weak
gap at $B \sim 24.1$ in NGC 2419 is real, and marks the boundary between the
canonical HB and blue hook stars. If true, this feature appears to be largely
metallicity-independent, and its visibility subject to the vagaries of the
processes distributing stars into the blue tail.

If the initial chemical compositions of the bluest HB stars match those in the
main HB peak, then we find that 25\% of stars reaching the HB are fainter than
this gap ($F814W > 24.4$) and may be blue hook stars. If they are more
helium-rich, then this fraction would be reduced somewhat because we would
have undercorrected for star lifetimes to get the distribution in the lower
panel of Fig. \ref{ihist}. The additional correction is very unlikely to
change the fact that this fraction is considerably larger than in any other
cluster with the possible exception of $\omega$ Cen.  We will return to this
subject in the conclusions.

\subsection{Ultraviolet Bright Stars}\label{uvsec}

The source of the UV upturn at wavelengths shorter than 2300 \AA ~ in spectra
of elliptical galaxies remains an unsolved problem in stellar populations. The
leading candidates for this emission are stellar types that have an
appropriate combination of high surface temperature and relatively long
lifetime. This includes extreme HB stars and their evolved forms, the AGB
manqu\'{e} stars; post-early AGB stars, which evolve to the AGB, but leave
again before having a thermal pulsation phase; and post-AGB stars, which are
short-lived but have high luminosity. The details of the evolution depend
largely on the mass of the envelope outside of the fusion shells. Recently
\citet{brown08} produced a UV CMD of the nearby elliptical galaxy M32, and
found that current evolutionary models predict many more UV-bright post-HB
stars than are observed.

NGC 2419 provides an interesting case study because of its large mass (which
helps to populate relatively short evolutionary stages) and because the HB
distribution can be described to first order with two strong peaks, as
discussed in \S \ref{HB}. Our simple expectation is that the peak just to the
blue of the instability strip is responsible for the majority of the AGB stars
observed in the cluster. However, an examination of the AGB population may
reveal whether the AGB stars evolve to a thermal pulsation termination, or
whether they leave ``early'' at a lower luminosity level on the AGB. Further,
a census of stars brighter than the HB and hotter than the AGB can help
observationally identify how the EHB stars in this cluster evolve. 

\begin{figure*}
  \includegraphics[scale=0.5,angle=-90]{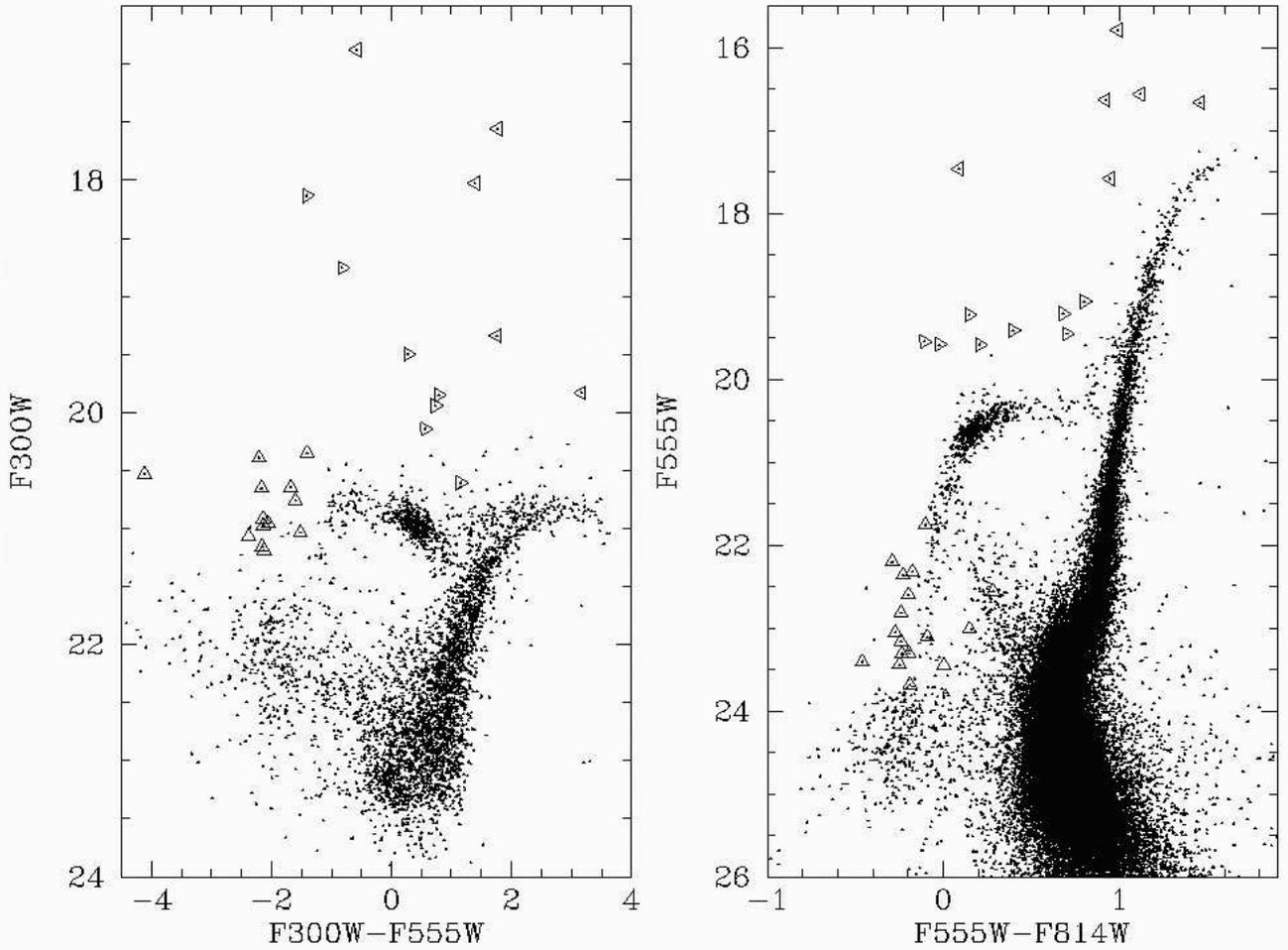}
 \caption{CMDs of ultraviolet-bright post-HB
    stars. The triangular symbols point in the expected direction of the
    evolution: upward for AGB manqu\'{e} stars, rightward for supra-HB stars
    evolving toward the AGB, and leftward for post-AGB and post-EAGB
    stars.\label{uvbright}}
\end{figure*}

We have therefore used a combination of optical and UV photometry to
identify probable hot post-HB stars. These are among the brightest objects
in the F300W filter, as can be seen in Fig. \ref{uvbright}.  We first
identified the two stars cataloged by \citet{zng} in their search for
stars that were bright in both the ultraviolet ($U$, which is similar
in bandpass to the F300W filter) and visible ($V$). ZNG 2 is more than
30 times brighter than the HB in the F300W filter, and ZNG 1 is
approximately 1.2 mag fainter than ZNG 2 and is brighter
in the optical. Both have optical photometry
consistent with being in a hot post-AGB phase. We identified five
other stars that sit closer to the AGB in the CMD. The star PA 4
has $F300W$ between ZNG 1 and 2 and is brighter in F555W than either
ZNG star. Two stars PA 6 and PA 7 did not have any UV photometry,
but occupy similar positions in the optical CMD. Star PA 5 is
faintest in the UV because it is the coolest star.  Star PA 3 is
significantly fainter in optical bands, and may be an example of a
post-early AGB star. Star PA 6 was saturated in the WFC images (the only
field in which it was observed), but the saturated photometry is accurate
enough to conclude that it is post AGB.



There are several reasonable evolutionary scenarios that could explain
the CMD positions of at least some of these stars (see Fig. 4 of
\citealt{brown} for examples of evolutionary tracks). As described
earlier, an RGB star that loses most of its envelope can leave the RGB
before having a core He flash, evolving rapidly toward higher
temperatures at nearly constant luminosity. Such stars can have a late
He flash that lands them on the blue end of the HB, or can become a He
core WD. Alternately, stars that had normal HB and AGB phases can
undergo H and He shell flashes that lead to large excursions in both
luminosity and surface temperature. These stars would be identified as
post-AGB (PAGB) or post-early AGB (PEAGB) stars. PAGB and PEAGB stars
move mostly in $T_{\rm eff}$ during later flashes, but can get several
times less luminous than the TRGB for shorter periods of time. Post AGB
evolution is more likely to explain most of these stars, but the
possibility exists that one of the more luminous stars might be a post
RGB star. We expect that spectroscopy would the easiest way of
identifying such stars.

\begin{figure*}
\includegraphics[scale=0.6]{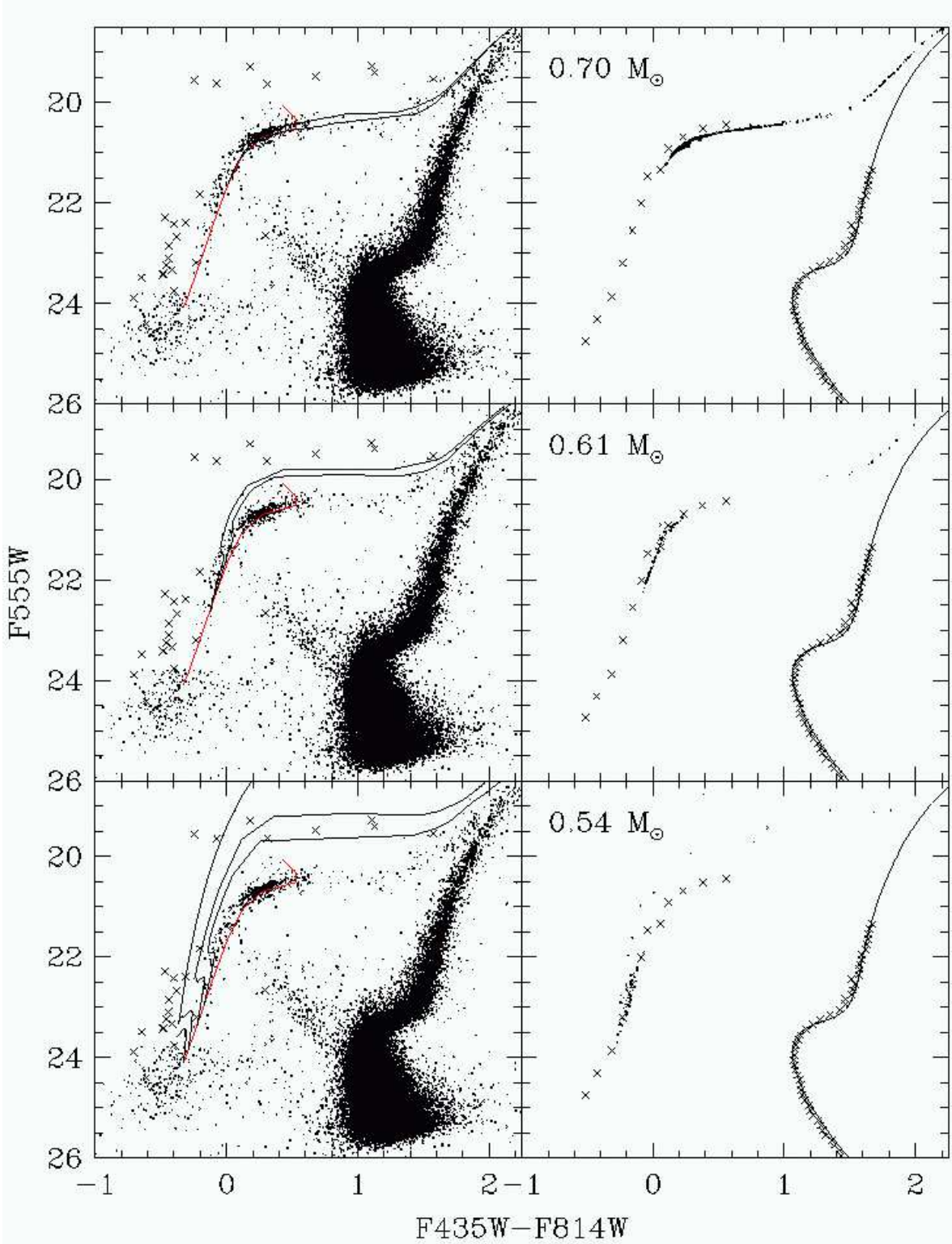}
\caption{Comparison of ACS WFC photometry with DSEP HB models. In all panels
  theretical values are shifted by 19.97 in F555W and 0.2 in F439W-F814 to
  account for distance and reddening. {\it Left panels:} Comparison with
  evolution tracks having [Fe/H]$=-2.0$ and [$\alpha$/Fe] $= +0.2$. The ZAHB
  is shown in red. The tracks are for 0.51, 0.52, 0.54, and 0.56
  $\msun$ ({\it bottom panel}), 0.58 and 0.6 $\msun$ ({\it middle panel}), and
  0.65 and 0.7 $\msun$ ({\it top panel}). {\it Right panels:} Examples of
  synthetic horizontal branch populations having [Fe/H] $=-2.1$, [$\alpha$/Fe]
  $= +0.2$, and a mass dispersion of 0.01 $\msun$, The number of synthetic HB
  stars roughly corresponds to the number in the incompleteness-corrected
  sample (see Fig. \ref{ihist}).  Also included ({\it solid line}) is an
  isochrone for the same composition and an age of 13 Gyr. Crosses show our
  mode fits to MS data, and individual stars that represent the mean HB line.
  \label{hbtracks}}
\end{figure*}

One of the clearer features associated with the evolution of the EHB stars is
a group of stars with $-0.5 \la F435W - F814W \la -0.3$ in the range $22 <
F435W < 23.4$ (see Figs. \ref{hbr60} or \ref{ehbblend}, for example).  We
have verified that all of the stars to the blue of the inter-gap HB population
($-0.2 \la F435W - F814W \la 0$) remain to the blue of the HB in all of
the UV CMDs. In the shortest wavelength filter (F250W), the stars remain a
nearly vertical sequence, implying luminosity evolution from the EHB rather
than slight temperature evolution from the inter-gap HB (see Fig. \ref{f250}).
The large size of the EHB population compared to the inter-gap population also
argues in favor of evolution from the EHB --- post-HB evolution is quite rapid
(on a Kelvin-Helmholtz timescale) in comparison to the HB phase itself, so
that there are likely to be many fewer post-HB stars.  The ratio of evolved HB
candidates to inter-gap HB stars is only about 1-to-2, whereas it is about
1-to-35 when compared to the EHB population. Theoretical ratios of post-HB
(helium shell fusion) to HB lifetimes are typically around 0.12
\citep{cass03}, and it is quite possible that not all of the EHB stars produce
post HB stars in this part of the CMD.  DSEP model tracks for [Fe/H] $= -2.0$
and [$\alpha$/Fe] $ = +0.2$ approximately parallel the line traced out by
these stars in optical bands (see Fig. \ref{hbtracks}). Their $0.52 \msun$
track comes close to reproducing the star colors, although there are no
corresponding HB stars on the slowest portions of the track (between the
zero-age HB and the portion shooting toward higher luminosity after core
helium exhaustion).  The gap of about 0.5 mag in the $F435W$ filter is
expected to correspond to the rapid evolution between core helium exhaustion
and the establishment of a stable He fusion shell. This may be useful as an
envelope helium abundance indicator {\it specifically} for the EHB stars
because the helium abundance affects the envelope mass for a given effective
temperature, and the envelope mass affects the star's luminosity when the He
shell is established \citep{brown08}. For more massive envelopes, the H fusion
shell can be restarted at the same time, and this contributes significantly to
the luminosity. At the moment, publicly-available models for EHB stars
(evolved past central He exhaustion) with helium enrichment do not exist, so
we are unable to calibrate this.  We encourage such modeling because it may
help answer questions related to the origins of multiple populations in
clusters and the possibility that it is caused by cluster self-enrichment.



A second group of stars with $F555W \approx 19.5$ also appears to be
composed of evolved HB stars. These appear to correspond to ``supra-HB'' stars
identified in clusters like M13 and discussed in \citet{ss} and \citet{zinn}.
The possibility that most of them are optical blends can be ruled out based on
their brightness in the ultraviolet. (One star that falls in this range was
identified as a blend in the HST photometry, however.)  Among these stars is
the bright RR Lyrae star V2, which was discussed in \S \ref{HB}.  We consulted
theoretical HB models \citep{dsep,cassshb} with very metal-poor compositions
and no helium enrichment, and verified that stars originating from the primary
peak in the HB distribution are expected to produce AGB stars exclusively ---
the evolution of the stars keeps them within about 0.1 mag of the HB until
central He exhaustion when they move to the AGB clump in about a
Kelvin-Helmholtz timescale.  Models agree that as stellar mass decreases the
post-HB evolution toward the AGB occurs brighter and brighter in $V$ until the
tracks never reach all the way to the canonical AGB. In a small range of
masses, stars can eventually evolve most of the way to the AGB, but spend a
significant amount of time (more than a Kelvin-Helmholtz timescale) brighter
than the HB at intermediate temperatures before reaching the AGB.  Such a
phase would correspond to the AGB clump if the stars had reached the canonical
AGB. It is also related to the ``bump'' on the first ascent giant branch ---
during this time the He fusion shell comes into equilibrium after consuming
the He abundance gradient left by the HB phase and adjusting to the new He
fuel concentration.

A big question is where on the HB the supra-HB stars originate from. According
to canonical models, the only evolution tracks that pass through this region
come from inter-gap HB stars. However, the relative numbers of inter-gap stars
and supra-HB stars appears to be inconsistent with this. Fig. \ref{hbtracks}
shows a comparison between our ACS WFC photometry and evolutionary tracks and
synthetic HB populations from DSEP models. [We used DSEP products using their
``empirical'' color transformations as these appear to do a better job than
the ``synthetic'' transformations in reproducing ACS data in and around the
subgiant branch \citep{dsep,sara}.] The numbers of stars in the synthetic model
representing the inter-gap population does not generate enough supra-HB stars
(especially on the hot side), although it may explain the reddest stars in
that group. Our best explanation for the number of evolved stars that are seen
is that both the supra-HB and AGB \manq ~ stars have evolved from the EHB peak,
with the supra-HB evolving from the cooler (brighter in optical filters) part
of the peak.


Increased envelope helium abundance tends to give stars a more pronounced blue
loop during their early HB evolution, and they tend to retain higher surface
temperatures during their subsequent evolution. (For examples, see
\citealt{swei}.) Helium enrichment can also lead to a convergence of post-HB
evolutionary tracks (see Fig. 9 of \citealt{busso} for an example) that could
be responsible for supra-HB stars. For NGC 2419, an increase in envelope
helium among the EHB stars might bring the models and observations into better
agreement, but the increase may not need to be large since the
theoretical tracks come close to matching the properties of the evolved HB
stars. Once again though, theoretical models for EHB stars with varying helium
abundance are need to settle the question.

\subsection{The Upper Red Giant Branch}\label{urgb}

Because NGC 2419 has a significant population of EHB stars, it is natural to
ask why these stars have been produced in such large numbers. In recent years,
there has been growing evidence that subpopulations of chemically-enriched
stars have been responsible for multimodal distributions of HB stars and for
the existence of stars at the high temperature end of the HB. However, there
has been no evidence of chemical composition variations within NGC 2419 from
the CMDs (no large dispersion along the evolutionary sequences as seen in
$\omega$ Cen, and no sign of multiple sequences), although it should be said,
there has also been little in the way of spectroscopic abundance studies.  As
a result, we examined the possibility that extreme mass loss could be
responsible for the bluest stars in the cluster. Helium enrichment within a
stellar population does encourage the formation of blue HB stars by allowing
lower mass stars to evolve off the main sequence and up the RGB. However,
helium enrichment does not change the theoretical requirement that
nearly all of the mass outside the helium core must be removed to put it on
the EHB.  If bright RGB stars deplete the mass stored in their envelopes
before reaching the TRGB, they can avoid a traditional helium flash event, and
instead may have a ``hot flash'' helium ignition that ultimately lands them on
the EHB (see Fig. 4 of \citealt{brown}, for example). Stars that can somehow
lose their envelopes before reaching within about 0.4 magnitudes of the TRGB
will not subsequently ignite helium, and are expected to produce He white
dwarfs \citep{dcruz,brown}.

Infrared excesses associated with giants in globular clusters provide
evidence of dust formation, possibly associated with strong mass
loss. \citet{47tuc} found excesses among a fraction of stars in the
metal-rich clusters 47 Tuc at luminosity levels down to the HB, and
becoming stronger with luminosity. The presence of excesses among only
a fraction of the stars implies episodic mass loss, although
\citet{omcen} present evidence that much of this may have been due to
blending in the infrared photometry. \citeauthor{omcen} find no
evidence of mass loss below the TRGB in $\omega$ Cen, which is a cluster
that is closer in composition to NGC 2419 that also has a large
population of EHB stars. So at this point there is not strong evidence
of strong mass loss below the TRGB.

\begin{figure}
  \includegraphics[scale=0.4]{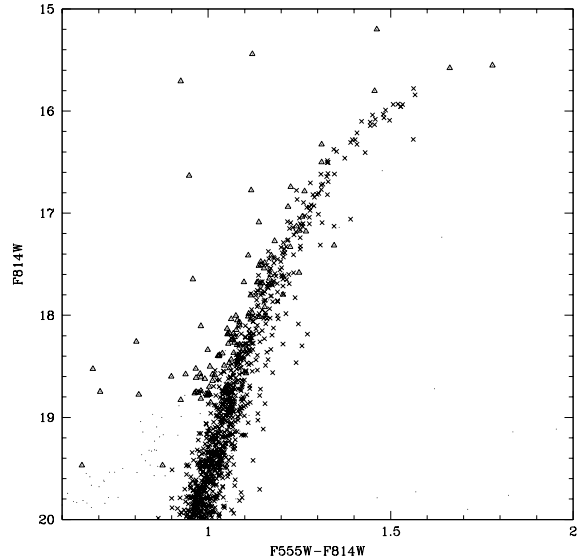}
 \caption{($F814W,F555W-F814W$) CMD for NGC 2419 stars on the bright RGB and
   AGB. The photometry is averages of measurements on the HST flight system
   after individual datasets were corrected for zeropoint differences relative
   to the proposal 7630 data. \label{rgbvi}}
\end{figure}

We used the cumulative luminosity function in our analysis, counting stars
starting at the observed TRGB. Comparison with models requires a magnitude
shift and a vertical normalization.  The horizontal magnitude shift was taken
from the magnitude difference between the brightest observed red giant of the
cluster and the TRGB in the theoretical model. By doing this, we virtually
eliminate all dependences on age and chemical composition for [Fe/H] $< -1$
\citep{sm07}. It is partly for these reasons that the luminosity of the TRGB
in $I$ has been used as a distance indicator for large, old metal-poor stellar
populations.  While there are giant stars that are brighter than the TRGB we
identify, there is a large gap in $F814W$ that leads us to identify them as
AGB stars, as shown in Fig. \ref{rgbvi}. The CMD also appears to be more
uniformly populated fainter than the third brightest giant star having $F814W
\sim 15.78$ that we identify with the TRGB.

\begin{figure}
  \includegraphics[scale=0.4]{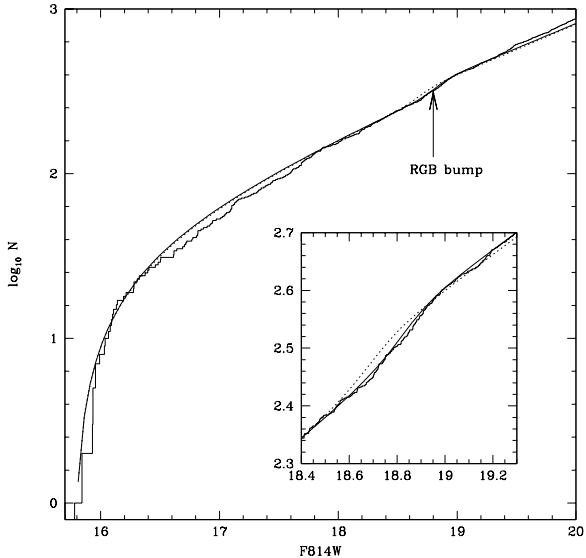}
 \caption{The cumulative luminosity function for
    bright RGB stars for NGC 2419, along with Victoria-Regina models
    \citep{vr} for age 12 Gyr and [Fe/H] $= -1.84$ ({\it thin solid line}) and
    $-2.01$ ({\it thin dotted line}) The inset shows a zoom centered on the
    position of the RGB bump.  \label{clf}}
\end{figure}

For the vertical normalization we forced the models to have the same number of
observed stars just brighter than the RGB bump at $F814W = 18.4$.  Shown in
Fig. \ref{clf} are the observed data matched with models having [Fe/H] of
$-1.84$ and $-2.01$. Although the [Fe/H] = $-1.84$ model does the best job in
matching the position of the RGB bump relative to the TRGB (see the inset in
Fig. \ref{clf}), in both cases the models predict a larger number of red
giants than the observed data has between $F814W \sim 16.4$ and $F814W
\sim 17.7$. As discussed by \citet{sm07}, the shape of the cumulative
luminosity function is a robust theoretical prediction.

The models were compared using the Kolmogorov-Smirnov (K-S) test.  This
evaluates the statistical probability that the null hypothesis (the two
distributions are drawn from the same distribution function) can be disproven.
The probability is based on the $D$ statistic, the maximum absolute difference
between the cumulative distributions of the two sets. Although it has the
advantage of making no assumptions about the precise form of the
distributions, the probability that is calculated depends on the size of the
sample. Because the choice of the cutoff magnitude for the comparison changes
the sample size, we ran the K-S test with different cutoffs (essentially
different normalization points).  Table \ref{clftab} shows the results of the
K-S tests, including the number of stars brighter than each magnitude cutoff.

\begin{deluxetable}{cllclc}
\tablewidth{0pt} 
\tablecaption{Cumulative Luminosity Function Comparisons for NGC 2419}
\tablehead{\colhead{$\Delta F814W$}  & \colhead{$N$} & \colhead{$D$} & \colhead{$P$} & \colhead{$D$} & \colhead{$P$}\label{clftab}}
\startdata
 & & \multicolumn{2}{c}{[Fe/H] = $-1.84$} & \multicolumn{2}{c}{[Fe/H] =
   $-2.01$}\\ 
1.0 & 41& 0.102 & 0.769 & 0.097 & 0.817 \\
1.2 & 54& 0.067 & 0.962 & 0.068 & 0.958 \\
1.4 & 72& 0.101 & 0.434 & 0.100 & 0.441 \\
1.6 & 85& 0.060 & 0.911 & 0.058 & 0.930 \\
1.8 &102& 0.048 & 0.970 & 0.046 & 0.981 \\
2.0 &131& 0.073 & 0.480 & 0.072 & 0.492 \\
2.2 &154& 0.055 & 0.721 & 0.054 & 0.744 \\
2.4 &181& 0.043 & 0.873 & 0.045 & 0.852 \\
\enddata
\tablecomments{Col. 1: magnitudes below TRGB for cutoff of RGB sample. Col. 2:
number of RGB stars in sample. Cols. 3, 5: absolute deviation between observed
and 
theoretical cumulative luminosity functions. Cols. 4, 6: Probability from K-S
test.} 
\end{deluxetable}

NGC 2419 only shows a slight deficiency of giants compared to the
theoretical models. The minimum K-S test probability is 16\% for a
sample with $F814W \approx 17.9$. 
The maximum difference between the observations and models (0.09 in
$\log_{10} N$) seen at $F814W = 16.6$ in Fig. \ref{clf} is
comparable to the fraction of EHB stars relative to all HB stars: a
33\% decrease in the giant population would only correspond to a
decrease in $\log_{10} N$ of about 0.12.  As a whole though, the
results are marginally consistent with a population of giant stars
having a distribution the same as the theoretical luminosity
function. The results appear to validate the algorithms for computing
plasma neutrino losses near the TRGB --- \citet{sm07} noted that newer
algorithms (e.g. \citealt{hrw,itoh}) produce LF shapes that are
consistent with observations. For typical RGB core conditions the
rapid increase in neutrino energy loss is strongly dependent on the
large temperatures and densities, and neutrino emission rapidly
increases as the TRGB is approached. The shape of the LF would be
changed if there were any energy loss mechanism having a significantly
different dependence on temperature and density.
The good agreement of the present NGC 2419 data with the models
strengthens the case that the plasma neutrino losses are now being
accurately modeled and non-standard energy loss mechanisms are not
operating in the core in a significant way.

The position of the modest deviation in NGC 2419's luminosity function
also shows up at a different luminosity than the one noted in NGC 2808
by \citet{sm07}. As noted earlier, stars within brightest 0.4 mag of
the luminosity function are expected to ultimately produce HB stars:
RGB stars having a normal helium flash at the TRGB will become normal
HB stars, while RGB stars leaving the RGB a little early can have a
late flash and become EHB stars. In NGC 2808, the brightest 1 mag in
F814W seemed to be fairly uniformly below predictions by about
20\%. In NGC 2419 the brightest 0.6 mag seems to be in good agreement
with the models, while there are noticeable deviations in the next 1.4
mag. Even taking into account the longer lifetimes of blue HB stars
having enhanced metal or helium abundances, the HB distribution
implies that more than 25\% of all HB stars being produced in NGC 2419
are becoming EHB stars (see \S \ref{HB}). In NGC 2808, the fraction of
HB stars that are on the EHB is significant, but quite a bit smaller
(see Fig. \ref{btcomp}, for example). Based on a simple
interpretation of the RGB luminosity function, we would have expected
NGC 2808 to have a large population of EHB stars, and NGC 2419 to have
a relatively unpopulated EHB but a potentially substantial population of He
white dwarfs. On the face of it, these results are somewhat
contradictory. What could be happening in these clusters?


A first possibility is that there is unaccounted-for contamination of
the RGB sample by AGB stars that masks a larger deficit of RGB stars
in NGC 2419. EHB stars are not expected to evolve into stars that
would be confused with RGB stars because they remain at high $T_{\rm
  eff}$ in their post-HB evolution, but the main group of HB stars
does produce AGB stars falling near the RGB.  The bright RGB is
particularly difficult to decontaminate because the AGB approaches the
RGB most closely near the TRGB, and this is also the part of the RGB
where contamination would have the largest effect on our ability to
detect signs of RGB stars leaving early. More accurate photometry or
spectroscopy would help to separate the AGB and RGB samples more
definitively. However, our AGB star sample is quite large in
comparison to the total HB population ($R_2^{\prime} = N_{AGB} /
N_{HB}^{\prime} = 0.146$), and especially in comparison to the HB
population in the main peak. So we believe we have identified the
great majority of the AGB stars.

If the luminosity function shape is correct, is it possible that production of
EHB stars does not require extreme mass loss? The primary suspect for
accomplishing this is through a cluster's self-enrichment of helium.  There is
quite a bit of evidence supporting the picture of chemical self-enrichment
within clusters, such as the sloping HB and unusually long periods of RR
Lyraes in NGC 6441 \citep{cadn6441}. There is also some evidence that chemical
self-enrichment can reduce the need for dispersion in red giant mass loss to
produce observed HB color distributions.  For example, in M3 (which admittedly
does not have EHB stars), \citet{cadm3} found that the combination of
substantial color spread on the HB and a sharply peaked period distribution
for RR Lyrae stars is most consistent with dispersion in helium abundance and
virtually no dispersion in mass loss ($\le 0.003 \msun$). However, unless
there is an extreme form of helium enrichment in the cluster, it would still
be necessary for stars to lose nearly all of their envelopes to reach the EHB.
An initial helium abundance $Y \sim 0.4$ would be required to reduce the mass
of stars at the TRGB to $\sim 0.5 \msun$ (e.g. \citealt{nt}). An enrichment of
this size should produce a clear second sequence in the CMD (remember that
such a sequence would also need to have to be about 25\% of the entire
population). In NGC 2808, \citet{p2808} identified a triple main sequence that
was consistent with populations having different helium abundances ranging
from primordial ($Y \approx 0.248$) up to 0.35 - 0.4.  $\omega$ Cen shows a
bluer main sequence \citep{bedomega} that is more metal-rich than the majority
of stars, and therefore is probably also helium enriched \citep{pomega}.

\begin{figure*}
\includegraphics[scale=0.8]{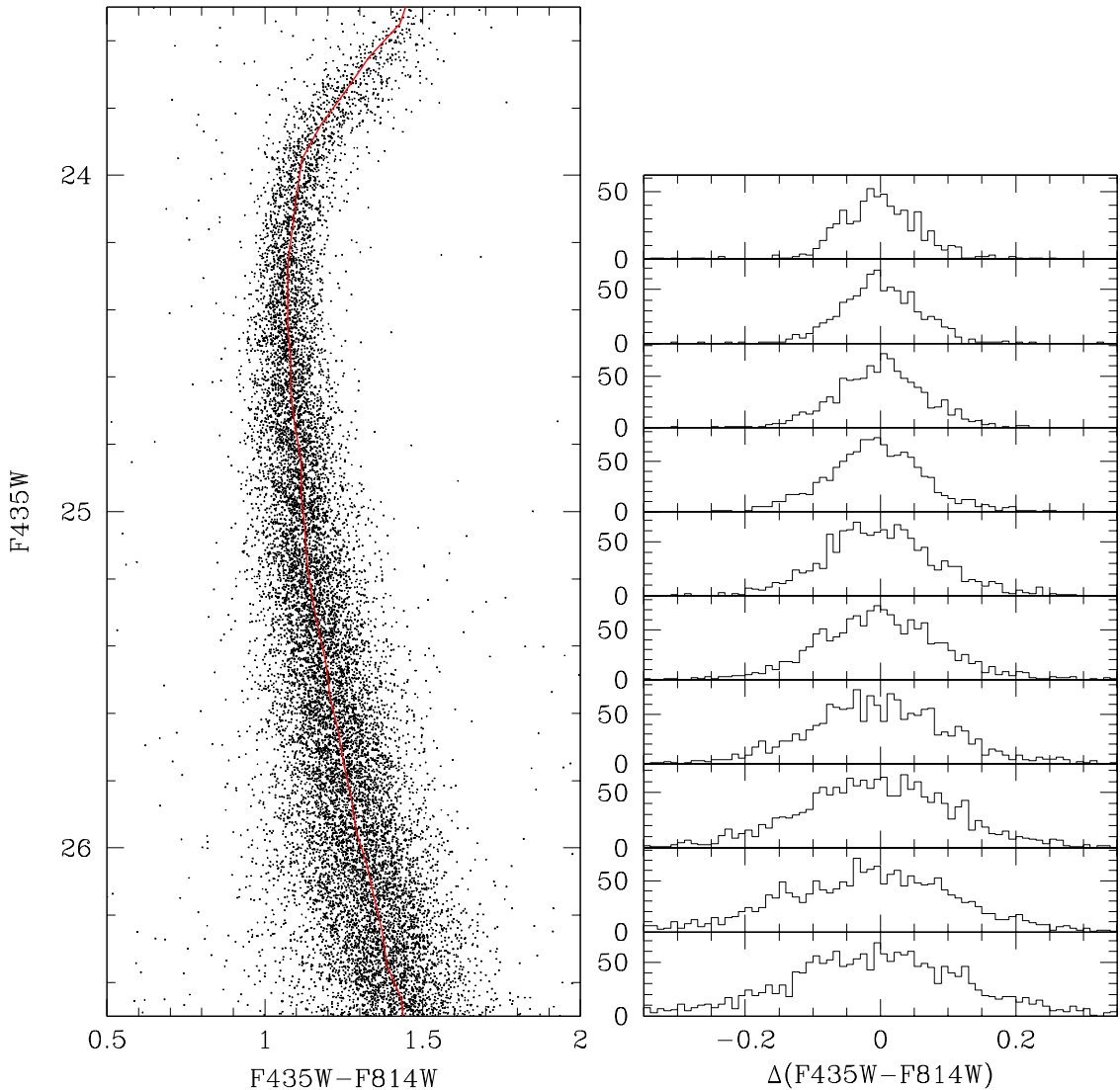}
\caption{ {\it Left panel:} The NGC 2419 main sequence in the WFC ACS dataset
  for stars with quality and radial position ($r < 90\arcsec$) cuts.  {\it
    Right panels:} Histograms of stars from the left panel after the color of
  the fiducial line has been subtracted. The total sample is separated into
  0.25 mag subsets.  \label{mshist}}
\end{figure*}

If similar helium variations were responsible for the two HB populations in
NGC 2419, a bimodal main sequence should be present in NGC 2419. In Fig.
\ref{mshist}, we show a zoom on the main sequence in a high-quality subset of
the ACS WFC photometry. A fiducial line was derived from stars in 0.1 mag
bins, and the main sequence was rectified by subtracting the color of the
fiducial line interpolated to the magnitude of each star. Histograms of the MS
star samples in 0.25 mag bins are shown on the right side of the figure. The
distributions resemble Gaussian distributions, and no feature appears
consistently from magnitude bin to bin. A large chemically-distinct population
can be ruled out with the present data unless there is a dysfunctional
combination of helium and metal enrichment that cancels any color shift.

Other factors affect the interpretation of present-day samples of RGB
and HB stars. For example, unless there is a physical process creating
a semi-permanent feature in the luminosity function, an observed
deficit of RGB stars (say associated with a statistical fluctuation)
would propagate up the RGB, and eventually result in a reduction in
the number of HB stars, although it should have no bearing on the
issue of the production of EHB stars. The current sample of HB stars
comes from stars that were on the RGB within about the last 130
Myr. According to models, a star moves through the uppermost 4.5 $I$
mag of the giant branch (almost 4 mag in $V$) in 100 Myr, which is a
typical lifetime for HB stars near the instability strip. For a
population with a uniform composition and no mass loss, the stars in
this range only vary in initial mass by about $0.002 \msun$ --- a tiny
range when translated back to the main sequence. In other words, the
rapidity of RGB evolution should stretch out statistical fluctuations
on the MS to the point of undetectability. The RGB cumulative
luminosity function is also insensitive to changes in the total number
of RGB stars in this range due to the way it is normalized. What
Fig. \ref{clf} examines is variations in the numbers of stars at
different luminosity levels within the sample.  It is interesting that
the two massive clusters with EHB stars that have been studied in this
way have both shown small deficits toward the bright end of the RGB
(and not excesses over theoretical predictions). However, the observed
deficits appear at different luminosity levels (implying that
different kinds of stars are being produced) and have marginal
significance. In addition, only one cluster (M5; \citealt{sm07})
without EHB stars has been discussed in a similar manner, although it
was consistent with theoretical expectations. We are forced to
conclude that the different methods of examining RGB mass loss
(detection of circumstellar dust and the luminosity function of the
upper RGB) do not provide convincing evidence of strong mass loss well
below the TRGB as yet.

\subsection{Radial Distributions}

\begin{figure*}
  \includegraphics[scale=0.5,angle=-90]{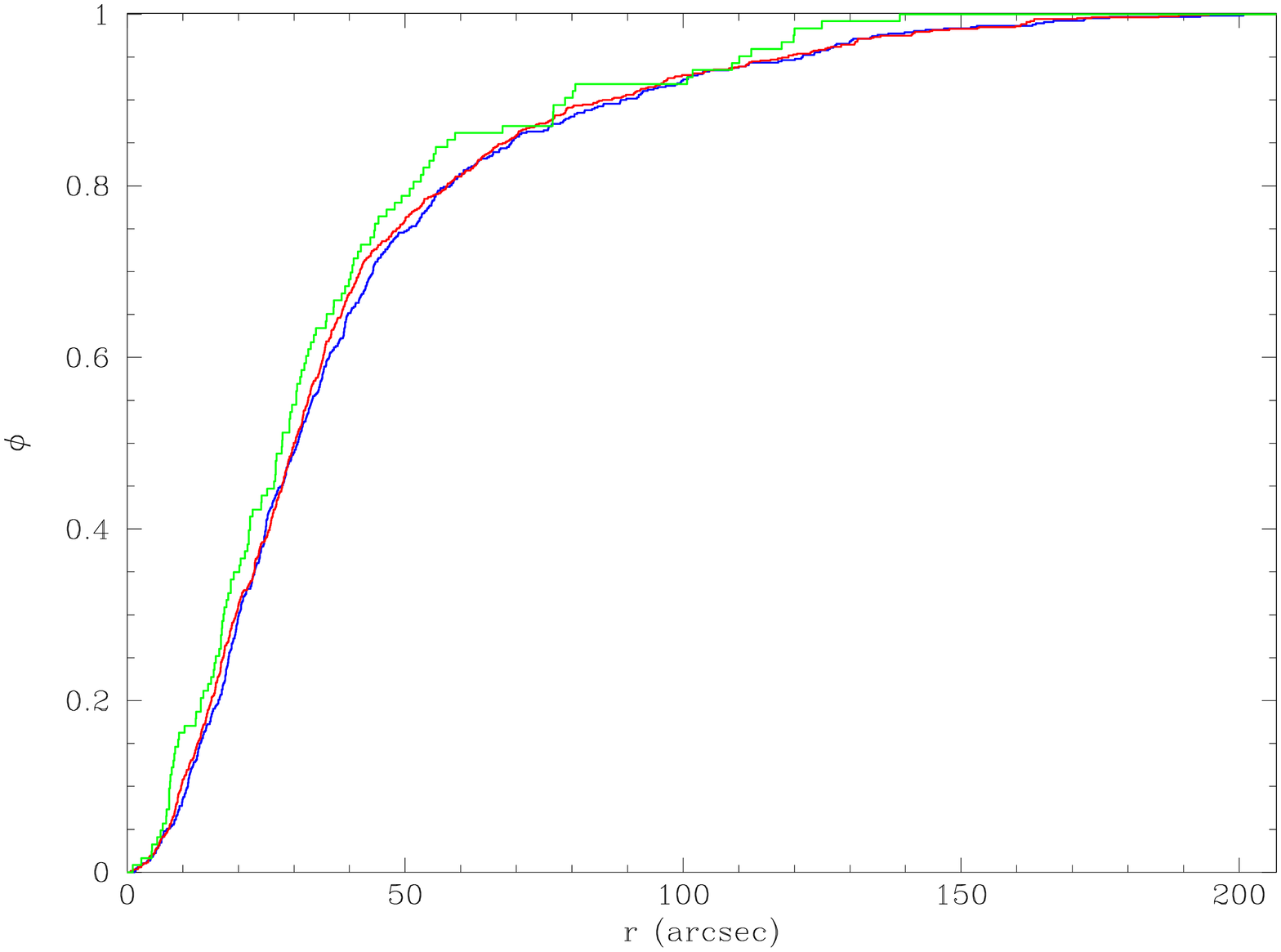}
 \caption{The cumulative radial distributions for 
    evolved stars in the combined HST fields. The red line shows 
    bright RGB stars ($F814W < 20$), green shows AGB stars, and blue shows
    HB stars excluding the EHB ($F814W < 23.75$) \label{crd}}
\end{figure*}

\citet{casomega} found trends in the radial distributions of stars in the
massive cluster $\omega$ Cen, including a radial variation in the percentage
of stars occupying the blue end of the HB. Structurally NGC 2419 and $\omega$
Cen are similar in concentration ($c = 1.67$ for NGC 2419, and 1.34 for
$\omega$ Cen; \citealt{mcl}), although NGC 2419 has unusually large core and
half-light radii ($r_c = 8.7$ pc and $r_h = 19.9$ pc) when compared to other
clusters of similar luminosity (4.1 pc and 7.2 pc for $\omega$ Cen, for
example).  However, when compared with other clusters of similar
galactocentric distance, its core and half-light radii seem typical, but its
total luminosity does not. We have a limited radial extent to examine here,
but we have compared the cumulative radial distributions for the core sample
in Fig.  \ref{crd}. A two-sample
Kolmogorov-Smirnov test returns a probability of 55\% that the RGB and bright
HB stars are drawn from the same distribution, and of 47\% for the RGB
and AGB samples. So the RGB, AGB, and bright HB distributions are in
reasonable agreement, with some small deviations of low significance.

\subsection{The $R$ Population Ratio}\label{R}


Because the population ratio $R = N_{HB} / N_{RGB}$ is a sensitive
indicator of helium abundance for clusters with large samples of
evolved stars, we will attempt an additional test of the hypothesis
that unusual CMD morphologies in the most massive clusters can be
attributed to chemical self-enrichment. 

The main difficulty in using the $R$ ratio is the calibration of the
values. Recent models \citep{sal} predict $R$ should have a value near
1.40 over a wide range of metallicities, but that the precise value
depends on the HB morphology. To make a reliable comparison of He
abundances in a {\it relative} sense, we must correct for variations
in lifetimes along the HB. This correction is important for a cluster
with a blue HB morphology like NGC 2419, and we discussed our method
in \S \ref{HB}.

\begin{figure}
  \includegraphics[scale=0.4]{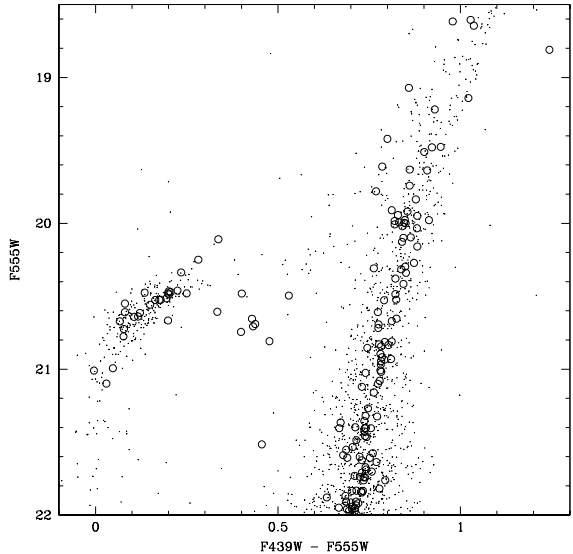}
 \caption{WFPC2 CMDs for NGC 2419 ({\it dots}) and
    M68 ({\it open circles}; \citealt{piotto}). The M68 dataset has been 
    shifted to match the blue HB of NGC 2419. \label{m68comp}}
\end{figure}

In order to compare NGC 2419 to other clusters, we follow a method
similar to that described by \citet{recio} to select the faint limit
for the RGB star sample. In their method, a template
cluster having an intermediate HB morphology is fit to the
cluster data and used to identify a reference HB level. Because NGC 2419 is a
very metal-poor cluster, we used M68 \citep{piotto} as the
template. The M68 photometry was shifted in magnitude and color to
achieve the best possible fit. (As discussed in \S \ref{HB} and below, the
red HB stars in NGC 2419 seem to exclusively be evolved HB stars, so
we looked for the best fits with stars in the main HB peak and bluer.)
The comparison between the proposal 7628 dataset (the only one that
used the F439W and F555W filters) and the \citet{piotto} dataset for
M68 is shown in Fig. \ref{m68comp} with the M68 photometry was shifted by
4.75 mag in $F555W$ and $0.05$ mag in $F439W-F555W$. The error in the
magnitude comparison appears to be about the same as the 0.07 mag
quoted by \citet{recio} in their work. Based on the reference ZAHB
value for M68 ($F555W^{ZAHB} = 15.73$; \citealt{recio}), this gives
$F555W^{ZAHB} = 20.48 \pm 0.08$ for NGC 2419.
Taking into account the zeropoint difference between the datasets for
proposals 7628 and 7630 ($-0.017$ mag in F555W), we have $F555W^{ZAHB} =
20.46 \pm 0.08$ on the reference system. The ZAHB level in F555W is used in
the definition of the faint end of the RGB sample in the studies of
\citet{zocc} and \citet{sal}, and we have used the same definition here for
the most direct comparison to other clusters.

With the zeropoint, we find $N_{RGB} = 529_{-31}^{+47}$ for stars in
the same fields as the HB stars (i.e. the ACS WFC fields of view). The
quoted error bars on $N_{RGB}$ come from the uncertainty in the ZAHB
level, which dominates Poisson statistics as a source of
error. Although NGC 2419 has a fairly large number of RR Lyrae stars
that could be used to establish the ZAHB level, indications are that
all of the RR Lyrae stars have significantly evolved away from the blue
HB (see \S \ref{HB}). \citet{rip} find $\langle V_{RR} \rangle = 20.31
\pm 0.01$ from 67 stars, but also find that the period-amplitude
distribution is most consistent with the ``well-evolved'' RR Lyrae
variables in M3. For the RR Lyrae stars in M68, \citet{m68} finds
$\langle V_{RR} \rangle = 15.64 \pm 0.01$. The M68 RR Lyrae stars have
positions in the period-amplitude diagram that are consistent with the
``regular'', relatively unevolved stars in M3.  The implied magnitude
shift between NGC 2419 and M68 from $\langle V_{RR} \rangle$
measurements is 4.67 mag. This is smaller than the value derived from
CMD comparison, and does not produce an adequate fit. This again is
consistent with the idea that the NGC 2419 RR Lyrae stars are evolved
and do not provide a good estimate of the ZAHB position, so we
therefore rely on the CMD fit value.

When combined with $N_{HB}^{\prime} = 821\pm29$ (the HB sample when corrected
for incompleteness and for lifetimes of blue HB stars relative to stars at the
edge of the instability strip), we find $R = 1.55^{+0.09}_{-0.13} $. The error
quoted for $N_{HB}^{\prime}$ only includes Poisson errors, but the uncertainty
in $N_{RGB}$ deriving from the ZAHB magnitude level dominates the uncertainty
in $R$. For [Fe/H] $=-2.1$ and $Y = 0.245$, \citet{sal} find $R = 1.40$ from
theoretical models, and this value changes by at most 0.01 at [Fe/H]$=-1.8$.
This implies that the {\it average} $Y$ for NGC 2419 stars is higher than
primordial by about 0.016, but the errors indicate this is just above a
$1\sigma$ difference. This is in spite of at least two effects that probably
produce systematic biases toward larger helium abundances (see \S \ref{HB}).
(Without the correction for star lifetimes, the HB population would have been
$N_{HB} = 954$, giving $R = 1.81$ and implying an average helium abundance
enhancement of about 0.041.) Still, if the EHB population (about 31\% of the
corrected total) is the only part that is helium enriched, it could be
enhanced by about 0.05. Based on the evolved stars, there is only weak
evidence of helium enrichment.

As a final comparison, we mention the population ratios for $\omega$ Cen
computed by \citet{casomega}. Though they did not compute an $R$ value that
could be directly compared with ours, their comparisons of HB to RGB or MS
stars implied that the HB star sample was 30-40\% larger than predicted by
canonical evolution models. Their result was actually larger than would be
expected given that $\omega$ Cen appears to host helium-enriched populations.
Regardless of the reason for the discrepancy, it further underlines the weak
evidence of helium enrichment in NGC 2419.

\section{Conclusions}

We have examined archival HST imagery in the optical and ultraviolet to study
evolved stars in the core of the massive globular cluster NGC 2419.  An
important aspect of these populations is the clear separation of the
horizontal branch population into two groups: the primary peak just to the
blue of the instability strip and a secondary peak of comparable size at the
extreme end of the HB. More than 38\% of the HB stars in this cluster are in
the EHB peak, and if their initial compositions were the same, this means that
more than 33\% of its red giants have been diverted into this population (more
than 25\% into blue hook candidates). While our study only examined the core
population (and in $\omega$ Cen, for example, there is some evidence for a
decrease in the abundance of EHB stars with distance from the cluster center),
this is a feat that is even remotely matched by only {\it some} of the very
most massive clusters in the Galaxy.  For example, $\omega$ Cen has
approximately 27\% of its HB stars on the EHB \citep{dale} --- a slightly
smaller proportion than NGC 2419. M54 also has a modest population of EHB
stars \citep{rosem54,siem54}, but it is a significantly smaller fraction of
the total HB population (we estimate $\la 15$\% from the CMD in
\citealt{rosem54}). NGC 2808 also has a modest EHB population that is 12\% of
its total HB sample (\citealt{castel}, where we have included their EBT3 stars
and the HBp stars that are likely to be composed of blends of EBT3 and MS
stars). In the very metal-rich cluster NGC 6388 \citep{dale08}, only about
3.5\% of the HB stars are on the EHB (2\% as blue hook stars).  A few other
massive clusters (e.g. NGC 6441, M15) have a handful of EHB stars that compose
a few percent of their populations at most, and another massive cluster (47
Tuc) has no known EHB stars. $\omega$ Cen, M54, NGC 6388, NGC 6441, and 47 Tuc
are probably the only clusters that are more massive than NGC 2419
\citep{mcl}, while NGC 2808 appears to have a similar mass and M15 is
approximately 50\% less massive. So while large cluster mass certainly makes
the appearance of EHB stars more likely, another factor is influencing their
relative abundance. This factor is probably not {\it past} cluster mass ---
NGC 2419 is unlikely to have lost as much mass in its history as clusters like
$\omega$ Cen and M54 because its orbit keeps it far out in the Galactic halo,
minimizing the effects of Galactic tides and disk shocking. There is also no
evidence of populations of greatly different age in NGC 2419 based on a
superficial examination --- NGC 1851 shows two distinct subgiant branches that
could result from an age difference of about 1 Gyr \citep{milo}.

For the clusters
above, there is a hint that the fraction of EHB stars anticorrelates with
metallicity. M15 is about 0.1 dex more metal poor than NGC 2419, but also by
far the least massive of these clusters.  $\omega$ Cen has a large metallicity
spread ($-2.2 \la $ [Fe/H] $\la -0.7$), but the most common [Fe/H] $\sim
-1.75$ \citep{johnomega}. Of the clusters with moderate EHB populations, M54
has [Fe/H] $\sim -1.6$ \citep{brownm54} and NGC 2808 has [Fe/H] $\sim -1.15$
\citep{carn2808}. The remainder of the massive clusters (NGC 6388, NGC 6441,
and 47 Tuc) are very metal rich ([Fe/H] $> -0.7$). This crude correlation may
be related to well-known relation between [Fe/H] and HB morphology --- more
metal-poor clusters tend to have bluer HB stars.

Even if we compare NGC 2419 to its nearest rival in producing EHB stars
($\omega$ Cen), we find a lack of a clear lead.  $\omega$ Cen has an HB
morphology similar to NGC 2419, though $\omega$ Cen has diverted a greater
proportion of stars into the blue HB tail between the two peaks, which would
imply that multiple populations should overlap to a greater degree in $\omega$
Cen.  $\omega$ Cen shows very clear signs of multiple populations on the giant
branch and main sequence \citep{bedomega}, including a blue main sequence that
is more metal-rich than the majority of stars, and therefore is probably also
helium enriched \citep{pomega}. The number ratio of chemically enriched group
to the total MS star population is similar to the ratio of EHB stars to the
total HB population, and thus \citeauthor{pomega} suggested that the EHB stars
are the descendants of the enriched population.  In NGC 2419, there is no
evidence to date of multiple populations among red giants or main sequence
stars.  Although there is a little evidence consistent with slight helium
enrichment in the cluster (based on post-HB evolution and the $R$ population
ratio), there is no evidence that it is as strong as in $\omega$ Cen.  It is
potentially possible to hide multiple populations on the main sequence by
increasing metallicity and helium abundance in a way to cancel out their
opposing color shifts. If this is happening in NGC 2419, it would imply a
smaller helium enrichment ($\Delta Y \sim 0.03$) than is implied for $\omega$
Cen ($\Delta Y = 0.14$). Such a small helium enrichment would be consistent
with the indicators presented in this paper, but would be unable to explain
the two HB star populations. Spectroscopic observations of cluster giants
would still be needed to determine whether a metallicity spread is truly
present in the cluster.

In one sense NGC 2419 is like $\omega$ Cen: neither appears to have undergone
strong dynamical relaxation \citep{dale,ferromega}.  There is little or no
evidence of a correlation with concentration $c$ \citep{mcl} though. The three
massive clusters with the largest EHB fractions (NGC 2419, $\omega$ Cen, and
M54) are also largest in $R_h$, although 47 Tuc has similar structural
properties to M54 and has no EHB stars.

At this point, we must confess that while NGC 2419 seems to have less
complicated stellar populations overall than clusters like $\omega$ Cen and
NGC 2808, its relative simplicity helps to stymie attempts to draw out a
coherent picture of the source of multiple populations and extreme horizontal
branch stars.

\acknowledgements

We thank the anonymous referee for a careful reading of the manuscript, and
gratefully acknowledge financial support from the National Science Foundation
under grant AST 0507785 to E.L.S. and Michael Bolte.

\clearpage
\begin{landscape}
\begin{deluxetable}{rrrccccccccccl}
  \tablewidth{0pt} \tabletypesize{\scriptsize} 
\tablecaption{Evolved
    Stars in NGC 2419} \tablehead{\colhead{ID} & \colhead{$\delta$R.A.} &
    \colhead{$\delta$DEC} & \colhead{F300W} &
    \colhead{F555W} & \colhead{F814W} &
    \colhead{F250W} & \colhead{F336W} &
    \colhead{F380W} & \colhead{F435W} &
    \colhead{F439W} & \colhead{F450W} &
    \colhead{F675W} & \colhead{Notes}\\
    & \colhead{(arcsec)} & \colhead{(arcsec)} & \colhead{(mag)} &
    \colhead{(mag)} & \colhead{(mag)} &
    \colhead{(mag)} & \colhead{(mag)} &
    \colhead{(mag)} & \colhead{(mag)} &
    \colhead{(mag)} & \colhead{(mag)} &
    \colhead{(mag)} & \colhead{}\label{phottab}} \startdata
  \multicolumn{14}{c}{Post-AGB candidates}\\
  PA 1 &   7.800&   16.468&   16.876&   17.461&   17.371&        &         &         &    17.403&         &        &         &  ZNG2\\
  PA 2 &  30.141&  -28.501&   18.026&   16.632&   15.707&        &   17.390&   17.617&          &   17.397&  17.137&         &  ZNG1\\
  PA 3 &  26.780&  -14.389&   19.339&   17.580&   16.632&        &         &         &          &         &        &         &  \\
  PA 4 & -49.254& -100.918&   17.556&   15.789&   14.797&        &   16.576&   16.851&    16.426&   16.767&        &         &  \\
  PA 5 &  -8.697&  -53.664&   19.832&   16.662&   15.200&        &   18.842&   18.536&    17.791&   17.921&  17.593&         &  \\
  PA 6 &  76.915& -185.451&         &         &         &        &         &         &    17.594&         &        &         &  \\
  PA 7 & -92.211& -73.163&          &   16.563&   15.442&        &         &         &    17.828&         &        &         & \\ \hline
  \multicolumn{14}{c}{supra-HB stars} \\
  SH 1 &  32.980&  -41.293&   18.132&   19.545&   19.653&        &         &         &    19.414&         &        &         &  \\
  SH 2 & -33.302&   13.065&   18.755&   19.579&   19.600&        &         &         &    19.574&         &        &         &  \\
  SH 3 &  10.784&   13.315&   19.941&   19.209&   18.525&        &         &         &    19.661&         &        &         &  \\
  SH 4 &  -1.433&  -21.805&   20.141&   19.582&   19.374&  20.164&   19.961&   19.862&    19.716&   19.749&  19.763&   19.535&  \\
  SH 5 &   5.825&   -5.158&         &   19.411&   19.013&        &         &         &    19.727&         &        &         &  \\
  SH 6 &   5.765&   -5.179&   19.849&   19.062&   18.259&        &   19.450&   19.439&    20.146&   19.310&  19.217&   18.432&  \\
  SH 7 &  -7.038&  -23.101&   20.605&   19.452&   18.748&  21.439&   20.046&   20.272&    19.803&   20.190&  20.077&   19.206& V2\\ \hline
  \multicolumn{14}{c}{AGB manqu\'{e} candidates} \\
  AM 1 & -63.325&  -13.723&         &   22.200&   22.490&        &         &         &    22.058&         &        &         &  \\
  AM 2 &  11.255&   -1.984&   20.762&   22.363&   22.590&        &         &         &    22.217&         &        &         &  \\
  AM 3 & -20.198&  -14.353&   20.645&   22.326&   22.501&        &   20.709&   21.975&    22.217&   22.310&  22.336&   22.466&  \\
  AM 4 &  25.687&  -57.545&   20.390&   22.599&   22.796&        &         &         &    22.469&         &        &         &  \\
  AM 5 &  52.320&  -55.610&   20.649&   22.819&   23.059&        &         &         &    22.619&         &        &         &  \\
  AM 6 &  43.226&  -65.388&   20.914&   23.055&   23.327&        &         &         &    22.884&         &        &         &  \\
  AM 7 &   5.899&  -27.397&         &   23.163&   23.403&        &         &         &    22.981&         &        &         &  \\
  AM 8 &   1.987&   -3.072&   20.954&   23.010&   22.860&  20.551&   21.101&   22.455&    23.433&   22.787&  22.601&   22.445&  blends with BSS in WFPC2\\
  AM 9 & -40.218&   -0.963&   21.156&   23.312&   23.547&        &   21.271&   22.809&    23.126&   23.184&  23.130&   23.400&  \\
  AM 10 & -32.261&  -16.589&   21.187&   23.308&   23.502&        &   21.378&   22.657&    23.174&   22.955&  22.996&   23.413&  \\
  AM 11 &  20.290& -122.984&         &   23.441&   23.689&        &   21.477&   22.890&    23.206&   23.162&  23.140&         &  \\
  AM 12 &  -9.226&    1.132&         &   23.407&   23.867&  20.985&         &         &    23.256&         &        &         &  \\
  AM 13 &  -8.234&  -14.431&   21.036&   22.554&   22.269&  20.734&   21.103&   22.515&    22.685&   22.594&  22.694&   22.165&  looks BSS in WFC, optical WFPC2\\
 AM 14 & -11.461&    2.148&         &   23.679&   23.867&  21.198&         &         &    23.505&         &        &         &  \\
 AM 15 &  -2.753&    2.081&   21.066&   23.448&   23.444&  20.644&   21.226&   22.892&    24.173&   22.822&  22.885&   22.662&  almost blended with BSS in WFC\\
  AM 16 &   3.907&  -38.774&   20.348&   21.752&   21.853&        &   20.301&   21.457&    21.686&   21.640&  21.704&   21.822&  \\
  AM 17 &  25.687&  -57.545&   20.390&   22.599&   22.796&        &         &         &    22.469&         &        &         &  \\
  AM 18 &  -9.150&   44.548&   20.530&   24.649&   24.042&        &         &         &          &         &        &         &  \\
  AM 19 &  -3.935&  -13.644&   20.971&   23.103&   23.193&  20.628&   21.059&   23.047&    22.996&   23.016&  22.948&   23.261&  \\
  AM 20 & 3.324& 6.803& 19.497& 19.224& 19.077& & & & 19.305& & & & \\ 
\enddata \tablecomments{Table \ref{phottab} is
    available (along with photometric errors) in its entirety via the
    link to the machine-readable version above. PA: Post-AGB; SH: Supra-HB;
    AM: AGB \manq; V: RR Lyrae variable; A: AGB; H: HB; R: RGB. The table only
  includes stars from the union of the ACS WFC and WFPC2 fields for proposals
  7628 and 7630.}
\end{deluxetable}
\clearpage
\end{landscape}

\begin{thebibliography}{}
\bibitem[Bedin et al.(2004)]{bedomega} Bedin, L.~R., Piotto, G., Anderson, J.,
  Cassisi, S., King, I.~R., Momany, Y., \& Carraro, G.\ 2004, \apjl, 605, L125
\bibitem[Boyer et al.(2008)]{omcen}Boyer, M. L., McDonald, I., van Loon, J. T.,
Woodward, C. E., Gehrz, R. D., Evans, A., \& Dupree, A. K. 2008, \aj, 135, 1395
\bibitem[Brown et al.(2008)]{brown08}Brown, T. M., Smith, E.,
Ferguson, H. C., Sweigart, A. V., Kimble, R. A., \& Bowers, C. W. 2008, \apj,
accepted 
\bibitem[Brown et al.(1999)]{brownm54} Brown, J.~A., 
Wallerstein, G., \& Gonzalez, G.\ 1999, \aj, 118, 1245
\bibitem[Brown et al.(2001)]{brown} Brown, T.~M., Sweigart, 
A.~V., Lanz, T., Landsman, W.~B., \& Hubeny, I.\ 2001, \apj, 562, 368
\bibitem[Busso et al.(2007)]{busso} Busso, G., et al.\ 2007, \aap, 474, 105
\bibitem[Cacciari et al.(2005)]{m3} Cacciari, C., Corwin, 
T.~M., \& Carney, B.~W.\ 2005, \aj, 129, 267
\bibitem[Caloi \& D'Antona(2005)]{cadm13} Caloi, V., \& D'Antona, F.\ 2005,
\aap, 435, 987
\bibitem[Caloi 
\& D'Antona(2007)]{cadn6441} Caloi, V., \& D'Antona, F.\ 2007,
\aap, 463, 949
\bibitem[Caloi \& D'Antona(2008)]{cadm3} Caloi, V., \& D'Antona, F.\ 2008,
\apj, 673, 847 
\bibitem[Carretta(2006)]{carn2808} Carretta, E.\ 2006, \aj, 131, 
1766 
\bibitem[Carretta et al.(2007)]{cn6218}Carretta, E., et al. 2007,
  \aap, 464, 939
\bibitem[Cassisi et al.(2004)]{cassshb} Cassisi, S., Castellani, M., Caputo,
  F., \& Castellani, V.\ 2004, \aap, 426, 641 
\bibitem[Cassisi et al.(2003)]{cass03} Cassisi, S., Salaris, 
M., \& Irwin, A.~W.\ 2003, \apj, 588, 862
\bibitem[Castellani et al.(2006)]{castel} Castellani, V., Iannicola, G., Bono,
  G., Zoccali, M., Cassisi, S., \& Buonanno, R.\ 2006, \aap, 446, 569 
\bibitem[Castellani et al.(2007)]{casomega} Castellani, V., et 
al.\ 2007, \apj, 663, 1021
\bibitem[Catelan et al.(1998)]{gaps} Catelan, M., Borissova, 
J., Sweigart, A.~V., \& Spassova, N.\ 1998, \apj, 494, 265
\bibitem[D'Antona et al.(2005)]{dan2808}D'Antona, F., Bellazzini, M.,
  Caloi, V., Fusi Pecci, F., Galleti, S., \& Rood, R. T. 2005, \apj,
  631, 868
\bibitem[Dalessandro et al.(2008)]{dale08} Dalessandro, E., 
Lanzoni, B., Ferraro, F.~R., Rood, R.~T., Milone, A., Piotto, G., 
\& Valenti, E.\ 2008, \apj, 677, 1069
\bibitem[Dalessandro et al.(2008)]{dale} Dalessandro, E., 
Lanzoni, B., Ferraro, F.~R., Vespe, F., Bellazzini, M., 
\& Rood, R.~T.\ 2008, \apj, 681, 311
\bibitem[D'Cruz et al.(1996)]{dcruz} D'Cruz, N.~L., Dorman, 
B., Rood, R.~T., \& O'Connell, R.~W.\ 1996, \apj, 466, 359
\bibitem[Dolphin(2000)]{hstphot} Dolphin, A.~E.\ 2000, \pasp, 
112, 1383
\bibitem[Dotter et al.(2007)]{dsep} Dotter, A., Chaboyer, 
B., Jevremovi{\'c}, D., Baron, E., Ferguson, J.~W., Sarajedini, A., 
\& Anderson, J.\ 2007, \aj, 134, 376
\bibitem[Ferraro et al.(1999)]{ferr}Ferraro, F. R., Messineo, M., Fusi
  Pecci, F., de Palo, M. A., Straniero, O., Chieffi, A., \& Limongi,
  M. 1999, \aj, 118, 1738
\bibitem[Ferraro et al.(1998)]{ferr98} Ferraro, F.~R., Paltrinieri, B., Pecci,
  F.~F., Rood, R.~T., \& Dorman, B.\ 1998, \apj, 500, 311
\bibitem[Ferraro et al.(1997)]{ferr97} Ferraro, F.~R.,
  Paltrinieri, B., Fusi Pecci, F., Rood, R.~T., \& Dorman, B.\ 1997, \mnras,
  292, L45
\bibitem[Ferraro et al.(2006)]{ferromega} Ferraro, F.~R., 
Sollima, A., Rood, R.~T., Origlia, L., Pancino, E., 
\& Bellazzini, M.\ 2006, \apj, 638, 433
\bibitem[Gallart et al.(2005)]{gall05}Gallart, C., Zoccali, M., \&
Aparicio, A. 2005, \araa, 43, 387
\bibitem[Grundahl et al.(1999)]{ujump} Grundahl, F., Catelan, 
M., Landsman, W.~B., Stetson, P.~B., \& Andersen, M.~I.\ 1999, \apj, 524, 242
\bibitem[Haft et al.(1994)]{hrw} Haft, M., Raffelt, G., 
\& Weiss, A.\ 1994, \apj, 425, 222
\bibitem[Harris(1996)]{har96}Harris, W. E. 1996, \aj, 112, 1487
\bibitem[Harris(1997)]{har97}Harris, W. E. et al. 1997, \aj, 114, 1030
\bibitem[Holtzman et al.(1995)]{holtz} Holtzman, J.~A., 
Burrows, C.~J., Casertano, S., Hester, J.~J., Trauger, J.~T., Watson, 
A.~M., \& Worthey, G.\ 1995, \pasp, 107, 1065
\bibitem[Iben \& Rood(1970)]{ir} Iben, I.~J., \& Rood, R.~T.\ 1970, \apj, 161,
  587
\bibitem[Itoh et al.(1996)]{itoh} Itoh, N., Hayashi, H., 
Nishikawa, A., \& Kohyama, Y.\ 1996, \apjs, 102, 411
\bibitem[Johnson et al.(2008)]{johnomega} Johnson, C.~I., 
Pilachowski, C.~A., Simmerer, J., \& Schwenk, D.\ 2008, \apj, 681, 1505
\bibitem[McLaughlin \& van der Marel(2005)]{mcl} McLaughlin,
  D.~E., \& van der Marel, R.~P.\ 2005, \apjs, 161, 304
\bibitem[Milone et al.(2008)]{milo} Milone, A.~P., et al.\ 
2008, \apj, 673, 241
\bibitem[Moehler et al.(1997)]{moem15} Moehler, S., Heber, U., \& Durell,
  P.~R.\ 1997, \aap, 317, L83 
\bibitem[Moehler et al.(2004)]{moe2808} Moehler, S., Sweigart, A.~V.,
  Landsman, W.~B., Hammer, N.~J., \& Dreizler, S.\ 2004, \aap, 415, 313
\bibitem[Newsham \& Terndrup(2007)]{nt} Newsham, G., \& Terndrup, D.~M.\ 2007,
\apj, 664, 332
\bibitem[Origlia et al.(2007)]{47tuc}Origlia, L., Rood, R. T., Fabbri, S.,
Ferraro, F. R., Fusi Pecci, F., \& Rich, R. M. 2007, \apjl, 667, L85
\bibitem[Pinto \& Rosino(1977)]{prrrlyr} Pinto, G., \& Rosino, L.\ 1977,
  \aaps, 28, 427 
\bibitem[Piotto et  al.(2002)]{piotto} Piotto, G., et al.\ 2002, \aap, 391, 945
\bibitem[Piotto et al.(2005)]{pomega} Piotto, G., et al.\ 
2005, \apj, 621, 777
\bibitem[Piotto et al.(2007)]{p2808} Piotto, G., et al.\ 
2007, \apjl, 661, L53
\bibitem[Recio-Blanco et 
al.(2005)]{recio} Recio-Blanco, A., et al.\ 2005, \aap, 432, 851
\bibitem[Ripepi et al.(2007)]{rip}Ripepi, V. et al. 2007, \aj, 667, L61
\bibitem[Rosenberg et al.(2004)]{rosem54} Rosenberg, A., 
Recio-Blanco, A., \& Garc{\'{\i}}a-Mar{\'{\i}}n, M.\ 2004, \apj, 603, 135
\bibitem[Salaris et al.(2004)]{sal} Salaris, M., Riello, M., Cassisi,
  S., \& Piotto, G.\ 2004, \aap, 420, 911
\bibitem[Sandquist \& Bolte(2004)]{sb} Sandquist, E.~L., \& Bolte, M.\ 2004,
\apj, 611, 323
\bibitem[Sandquist \& Martel(2007)]{sm07} Sandquist, E.~L., \&
  Martel, A.~R.\ 2007, \apjl, 654, L65 
\bibitem[Sarajedini et al.(2007)]{sara} Sarajedini, A., et 
al.\ 2007, \aj, 133, 1658 
\bibitem[Siegel et al.(2007)]{siem54} Siegel, M.~H., et al.\ 
2007, \apjl, 667, L57
\bibitem[Sirianni et al.(2005)]{sir} Sirianni, M., et al.\ 
2005, \pasp, 117, 1049
\bibitem[Stetson(1996)]{stetj} Stetson, P.~B.\ 1996, \pasp, 108, 851
\bibitem[Stetson(2005)]{stetn2419} Stetson, P.~B.\ 2005, \pasp, 
117, 563 
\bibitem[Strom et al.(1970)]{ss} Strom, S.~E., Strom, K.~M., Rood, R.~T., \&
Iben, I., Jr.\ 1970, \aap, 8, 243
\bibitem[Suntzeff et al.(1988)]{sunt}Suntzeff, N. B, Kraft, R. P., \&
Kinman, T. D. 1988, \aj, 95, 91
\bibitem[Sweigart(1987)]{swei} Sweigart, A.~V.\ 1987, \apjs, 65, 95
\bibitem[VandenBerg et al.(2006)]{vr} VandenBerg, D.~A., 
Bergbusch, P.~A., \& Dowler, P.~D.\ 2006, \apjs, 162, 375
\bibitem[Walker(1994)]{m68} Walker, A.~R.\ 1994, \aj, 108, 555
\bibitem[Zinn et al.(1972)]{zng} Zinn, R.~J., Newell, E.~B., \& Gibson, J.~B.\
1972, \aap, 18, 390
\bibitem[Zinn(1974)]{zinn} Zinn, R.\ 1974, \apj, 193, 593
\bibitem[Zoccali et al.(2000)]{zocc} Zoccali, M., Cassisi, S., Bono, G.,
  Piotto, G., Rich, R.~M., \& Djorgovski, S.~G.\ 2000, \apj, 538, 289
\end{thebibliography}
\end{document}